\documentclass[twocolumn,showpacs,preprintnumbers,amsmath,amssymb,floatfix]{revtex4}

\pdfoutput=1
\usepackage{graphicx}
\usepackage{dcolumn}
\usepackage{bm}
\usepackage{url}
\usepackage{amsmath}
\usepackage{verbatim}  

\newcommand{\positron}{\mathrm{e}^{+}}
\newcommand{\FLUKA}{\textsc{fluka}}
\newcommand{\MADX}{\textsc{mad-x}}
\newcommand{\electron}{\mathrm{e}^{-}}
\newcommand{\sigbfpp}{\sigma_{\scriptscriptstyle \mathrm{BFPP}}}
\newcommand{\pbone}{$^{208}\mathrm{Pb}^{81+}$}
\newcommand{\pb}{$^{208}\mathrm{Pb}^{82+}$}
\newcommand{\epem}{\positron\electron}
\newcommand{\Brho}{(B\!\rho)}
\newcommand{\dBFPP}{\delta_\mathrm{BFPP}}

\newcommand{\dEMDone}{\delta_{\mathrm{EMD1}}}
\newcommand{\dEMDtwo}{\delta_{\mathrm{EMD2}}}
\newcommand{\lum}{\mathcal{L}}
\newcommand{\xbfpp}{x_\mathrm{\small{B}}}
\newcommand{\xbfppI}{x_{\mathrm{\small{B}},1}}

\newcommand{\Pbfpp}{\lambda}
\newcommand{\Pdelta}{f_{\delta p}}
\newcommand{\StandDevBfpp}{\sigma_{\lambda,0}}
\newcommand{\StandDevBfppOne}{\sigma_{\lambda,1}}
\newcommand{\atlas}{\textsc{atlas}}
\newcommand{\alice}{\textsc{alice}}
\newcommand{\cms}{\textsc{cms}}
\newcommand{\mm}[1]{$#1\,\mathrm{mm}$}
\DeclareMathOperator{\erf}{erf}

\begin{document}


\title{Beam losses from ultra-peripheral nuclear collisions\\ between \pb\ ions in the Large Hadron Collider and their alleviation}

\author{R.~Bruce}
 \altaffiliation[Also at ]{MAXlab, Lund University, Sweden.}
 \email{roderik.bruce@cern.ch}
\author{S.~Gilardoni}
\author{J.M.~Jowett}
\affiliation{CERN, Geneva, Switzerland}
\author{D.~Bocian}
\altaffiliation[Also at ]{CERN, Geneva, Switzerland and IFJ PAN, Krakow, Poland.
This work was supported in part by U.S. Dept. of Energy under contract No. DE-AC-02-07CH11359.}
\affiliation{Fermi National Accelerator Laboratory, Batavia IL 60510 USA}

\date{\today}

\begin{abstract}

Electromagnetic interactions between colliding heavy ions at the Large Hadron Collider (LHC) at CERN will give rise to localized beam losses that may quench superconducting magnets, apart from contributing significantly to the luminosity decay. To quantify their impact on the operation of the collider, we have used a three-step simulation approach, which consists of optical tracking, a Monte-Carlo shower simulation and a thermal network model of the heat flow inside a magnet. We present simulation results for the case of \pb\ ion operation in the LHC, with focus on the \alice\ interaction region, and show that the expected heat load during nominal \pb\ operation is 40\% above the quench level. This limits the maximum achievable luminosity. Furthermore, we discuss methods of monitoring the losses and possible ways to alleviate their effect.

\end{abstract}


\maketitle

\section{INTRODUCTION}

The Large Hadron Collider (LHC) is presently being commissioned at CERN~\cite{lhcdesignV1}. Its main beam parameters are given in Table~\ref{tab:lhc-beams}. In its first phase of operation it will collide protons but later also heavy nuclei, starting with \pb\ at center-of-mass energies up to 1.15~PeV~\cite{jowett08}.  This will open up a new energy regime in experimental nuclear physics, extending the study of the hadronic matter (``quark-gluon plasma'') that existed in the early universe about $10^{-6}\,\mathrm{s}$ after the Big Bang~\cite{alice}. At the same time, colliding heavy-ion beams at this unprecedented energy will present beam physics challenges not encountered in previous colliders.

When two fully stripped ions collide at an interaction point (IP), a variety of processes leading to fragmentation and particle production can occur. Hadronic nuclear interactions, which are usually the main object of study of the experiments,     occur only when the impact parameter $b$ is smaller than about twice the nuclear radius $R$. \emph{Ultraperipheral collisions} are those in which two colliding ions pass close to each other  with $b>2R$. In such events, the intense Lorentz-contracted fields of the nuclei can be represented as pulses of virtual photons in the equivalent photon picture of Fermi, Weizs\"acker and Williams~\cite{Jackson}. These extremely energetic photons collide and cause electromagnetic interactions that are particularly strong in heavy-ion collisions because the density of virtual photons around the nuclei is proportional to $Z^2$. Reviews of this field can be found in Refs.~\cite{baltz08,baur98,bert02,bert05}. In comparison with the cross-section for inelastic hadronic interactions, $\sigma_{h}\approx 8\,\mathrm{b}$, the rates of ultraperipheral interactions are enormous: for example, the cross-section for $\epem$ pair production given by the Racah formula~\cite{racah} is of order $2\times10^{5}\,\mathrm{b}$. From the point of view of collider operation, the  most important electromagnetic processes  are the  sub-classes of these interactions which remove ions from the beam, namely bound-free pair production (BFPP) and electromagnetic dissociation (EMD)~\cite{baltz96}.

\begin{table} \centering
  \caption{Design parameters for the LHC's proton and \pb\   beams in
           collision conditions~\cite{lhcdesignV1}. The values of $\lum$ and $\beta^*$ refer to IP2 for \pb ions and IP1 and IP5 for protons (see Fig.~\ref{fig:LHC-layout}).}
  \label{tab:lhc-beams}
  \begin{ruledtabular}
\begin{tabular}{|l|r|r|}
  Particle & p & \pb \\ \hline \hline
  Energy/nucleon & 7 TeV & 2.759 TeV \\ \hline
  No.\ of  bunches $k_\mathrm{b}$ & 2808 & 592 \\ \hline
  Particles/bunch & 1.15$\times10^{11}$ & $7\times10^7$ \\
  \hline
  Transv. normalized & & \\
  emittance ($1\sigma$)& $3.75\,\mathrm{\mu}m$ & $1.5\,\mathrm{\mu}m$\\
  \hline
  RMS momentum & & \\
  spread $\langle \delta_p^2 \rangle^{1/2}$ & $1.13\times 10^{-4}$ & $1.10\times 10^{-4}$ \\
  \hline
  Stored energy per beam & 362 MJ & 3.81 MJ \\
  \hline
  Design luminosity $\lum$ & $10^{34}\mathrm{cm}^{-2}\mathrm{s}^{-1}$ &
                      $10^{27}\mathrm{cm}^{-2}\mathrm{s}^{-1}$\\
  \hline
  horizontal and vertical $\beta^*$ & 0.55 m & 0.5 m
\end{tabular}
\end{ruledtabular}
\end{table}

In BFPP, the virtual photons surrounding relativistic ions collide and produce an $\epem$ pair where, in contrast to the much more frequent free pair production, the electron is created in an atomic shell of one of the ions. Schematically, the reaction between two bare nuclei with atomic numbers $Z_1$, $Z_2$ can be written as
\begin{equation}
\label{eq:bfpp-reaction}
Z_1  + Z_2 \longrightarrow
(Z_1  + \mathrm{e}^-)_{1s_{1/2},...}  + Z_2  + \mathrm{e}^+.
\end{equation}

In EMD one nucleus absorbs a photon and undergoes a transition into an excited state, typically by excitation of the giant dipole resonance. When this decays, it emits one or several nucleons. Because of the nuclear Coulomb barrier, the most common processes are the emission of one or two neutrons.  For the case of \pb\ ions, the 1-neutron reaction (called EMD1 hereafter) can be written as
\begin{equation}
^{208}\mathrm{Pb}^{82+} + ^{208}\mathrm{Pb}^{82+}
\longrightarrow
^{208}\mathrm{Pb}^{82+} + ^{207}\mathrm{Pb}^{82+}+ \mathrm{n}.
\end{equation}

Both BFPP and EMD change the magnetic rigidity of at least one of the colliding nuclei. (as usual, magnetic rigidity is defined as a particle's momentum $p$ per unit charge $Ze$, or $p/Ze=\Brho$, where $\rho$ is the bending radius in a  magnetic field  $B$.) If the charge of an ion changes by $e\Delta  Z$, and the number of nucleons by $\Delta A$, the resulting rigidity can be written as $\Brho(1+\delta)$, where the fractional deviation $\delta$ from the main beam with atomic number $Z_0$ and mass number $A_0$ is given  to a very good approximation (neglecting 
increments of the mass excess) by
\begin{equation}
\label{eq:delta-rigidigy}
\delta \approx
\frac{Z_0(A_0+\Delta A)}{A_0(Z_0+\Delta Z)}\left(1+\delta_p\right)-1.
\end{equation}
Here $\delta_p=\Delta p/p_0$ is the fractional momentum deviation per nucleon with respect to an ion circulating on the central orbit of the storage ring.

During the interactions, the transverse momentum recoil is very small, so the modified ions emerge at a small angle to the main beam. In EMD, the recoil changes $\delta_p$ (see Sec.~\ref{sec:emd-quench}), while this is negligible for BFPP. However, ions from both processes follow dispersive orbits according to their magnetic rigidity and may be lost at the first point in the ring where the horizontal aperture $A_x$ and dispersion $d$ generated locally since the IP satisfy
\begin{equation}
\label{eq:loss-condition}
d\,\delta \ge A_x.
\end{equation}
The beam losses caused by these processes  contribute  significantly to the decay of intensity and luminosity in an ion collider~\cite{gould84,baltz96}. This  has been  evaluated for the LHC~\cite{lhcdesignV1,epac2004}, where ions might collide at three IPs: the \atlas\ experiment at IP1, \alice\ at IP2 and \cms\ at IP5.

\begin{figure}[tb]
  \begin{centering}
  \includegraphics[width=8cm]{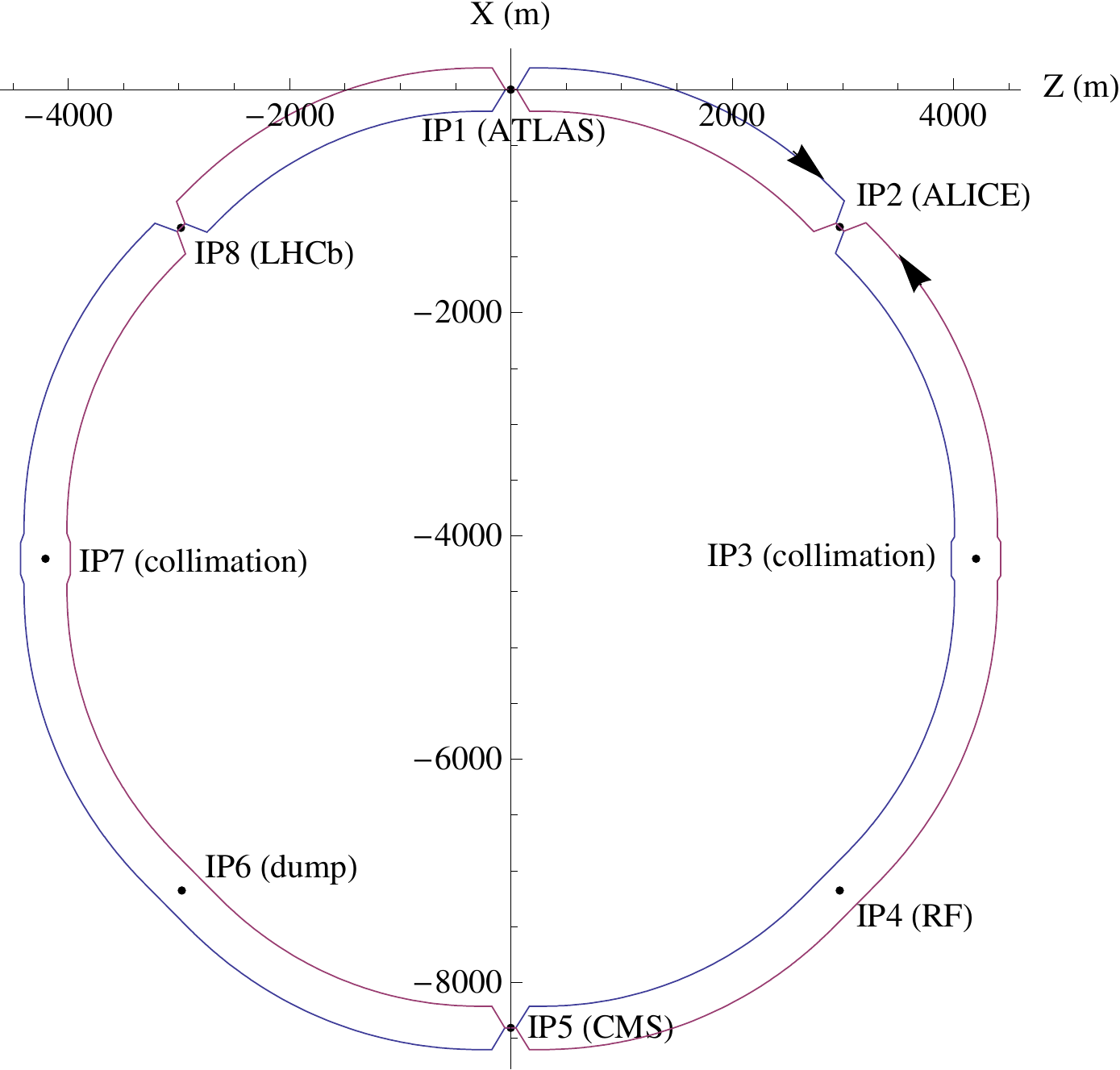}\\
  \end{centering}
  \caption{The schematic layout of the LHC
  (the separation of the two rings is exaggerated).}\label{fig:LHC-layout}
\end{figure}

The general layout of the LHC is shown in Fig.~\ref{fig:LHC-layout}. It consists of two concentric 27~km rings with counter rotating beams. Each ring is made of eight long straight sections matched to eight regular arcs through dispersion suppressor regions. The two rings overlap in four of the eight long straight sections housing colliding-beam experiments. The optics design continues to evolve through various \emph{versions}, usually corresponding to a given layout of the collider elements, with the latest one called V6.503. The LHC uses superconducting magnets with the two apertures within the same cryostat, where some are cooled down to 1.9~K by liquid helium. Further details of the LHC optics and layout can be found in Ref.~\cite{lhcdesignV1}.

The rate of removal of particles from the beam at the IPs is directly proportional to the interaction cross section, which in the case of BFPP takes the approximate form~\cite{meier01}
\begin{equation}
\label{eq:sigmabfpp}
\sigbfpp \approx Z_1^5  Z_2^2  \sum_i{\left(A_i  \log\gamma_{\mathrm{cm}} + B_i\right)},
\end{equation}
where the electron is captured by nucleus~1 and the sum is taken over atomic shells $i$. Here $\gamma_{\mathrm{cm}}$ is the relativistic factor of the ions in the center-of-mass frame, and $A_i,B_i$ depend only weakly on $Z$. For 2.76~TeV/nucleon \pb collisions in the LHC, the total cross section for electron capture to \emph{one} of the nuclei is~\cite{meier01}
\begin{equation}
\label{eq:sigBFPP-LHC}
\sigbfpp\approx 281\: \mathrm{b}.
\end{equation}
Using Tab.~\ref{tab:lhc-beams} and Eq.~(\ref{eq:sigBFPP-LHC}), the BFPP event rate is $\lum \sigbfpp\approx 281\,\mathrm{kHz}$ ($\lum$ is the collider luminosity).

Predictions of the cross-sections for BFPP have varied substantially over the years but have converged on values close to those found using the plane-wave Born approximation of Ref.~\cite{meier01}; as these authors point out, it was shown earlier~\cite{baltz97}, that higher order Coulomb corrections may reduce the values by a few percent. We estimate the uncertainty in the BFPP cross section to about 20\%~\cite{meier01}.

The corresponding cross section for EMD1 was calculated with the Monte Carlo program \FLUKA~\cite{fluka1,fluka2008,roes00}, by simulating a large number of the events through direct calls to the event generator:
\begin{equation}
\label{eq:sigEMD1-LHC}
\sigma_{\scriptscriptstyle \mathrm{EMD1}}\approx 96\: \mathrm{b}
\end{equation}
This relies on a recent improved implementation of EMD effects which has been benchmarked against the \textsc{reldis} code~\cite{pshen01} (e.g., in this case \textsc{reldis} gives 104~b~\cite{lhcdesignV1}). \FLUKA\ was also used to estimate the cross section for 2-neutron EMD (called EMD2) to
\begin{equation}
\sigma_{\scriptscriptstyle \mathrm{EMD2}}\approx 29 \: \mathrm{b},
\end{equation}
while the total EMD cross section, including all decay channels, is
\begin{equation}
\label{eq:sigEMD2-LHC}
\sigma_{\scriptscriptstyle \mathrm{EMD}}\approx 226 \mathrm{b}.
\end{equation}

Thus, the total cross section for particle loss by electromagnetic processes is around 507~b, compared to the 8~b for nuclear inelastic interactions, and is the dominant limit on usable luminosity lifetime for the experiments.

Furthermore, if Eq.~(\ref{eq:loss-condition}) is satisfied somewhere,  then the lost particles hit the vacuum chamber in a well-defined location which may lie in a superconducting magnetic element~\cite{klein01,pac2003,epac2004}. With the energy $E$ per ion in Tab.~\ref{tab:lhc-beams}, the total power in the BFPP beam is $\lum \sigbfpp E\approx 25.8 \,\mathrm{W}$. This induced heating may raise the temperature enough to bring the superconducting cable irreversibly over the critical surface in its phase diagram, the space spanned by magnetic field, current, and temperature. The resulting departure from the superconducting state is called a quench. The induced Joule heating also quenches neighboring volumes so that the quench propagates. In case of a quench in the LHC, the beam will be dumped within one machine turn and the quench protection system will fire heaters to quickly quench a number of magnets  and dissipate the stored magnetic energy over a larger volume, avoiding damage to the magnets. Nevertheless, quenches have to be avoided by all means during collider operation since recovery involves the lengthy process of cooling the magnets down  again, followed by refilling, ramping and ``squeezing'' of the beams. Downtime of the LHC is costly.

BFPP has been  measured in fixed target experiments~\cite{belkacem93,krause98,grafs99}. The first measurement in a collider~\cite{prl07} occurred during 100~GeV/nucleon $^{63}$Cu$^{29+}$ operation of the Relativistic Heavy Ion Collider (RHIC) at Brookhaven National Laboratory, where the detected induced showers agreed within a factor~2 with FLUKA simulations and the location of the losses was predicted within 2~m. Although neither BFPP nor EMD pose any risk of quenches at RHIC, mutual EMD between ions of the colliding beams has also been measured at RHIC~\cite{chiu02} and is routinely used to monitor the luminosity~\cite{baltz98,adler01}.

\begin{figure*}[tb]
  \begin{centering}
  \includegraphics[width=16cm]{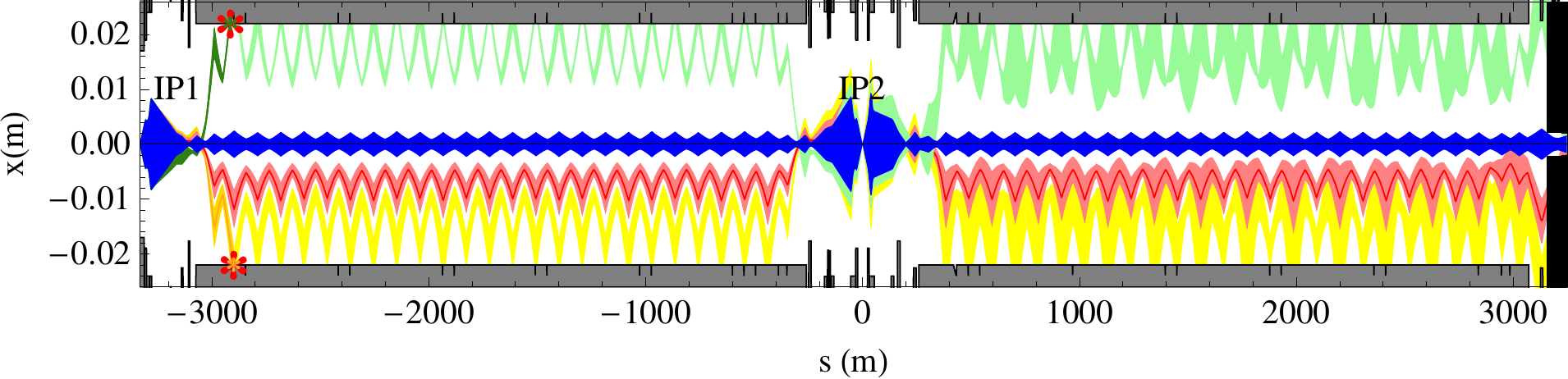}\\
  \includegraphics[width=16cm]{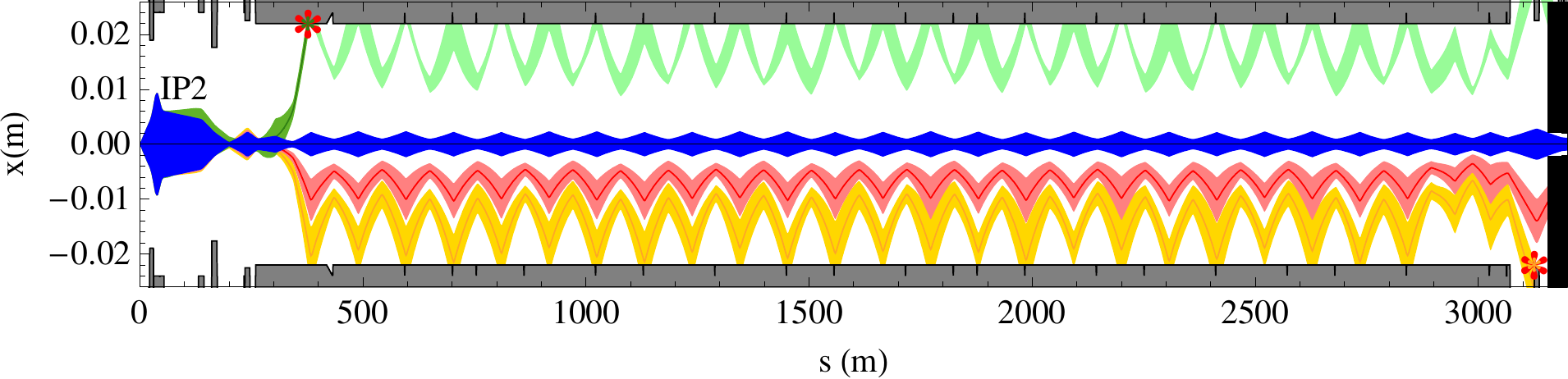}\\
  \includegraphics[width=16cm]{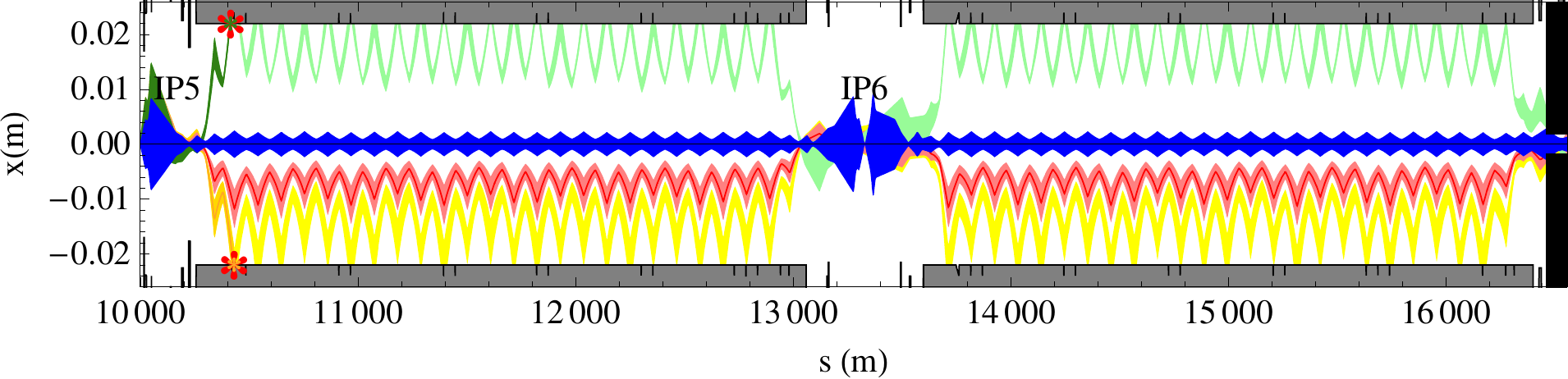}\\
  \end{centering}
  \caption{The simulated horizontal $6\sigma$ envelopes of the nominal \pb\ beam
      (blue), the \pbone\ BFPP beam (green), the $^{207}\mathrm{Pb}^{82+}$
      EMD1 beam (red) and the $^{206}\mathrm{Pb}^{82+}$ EMD2 beam (yellow)
      coming out of IP1 (top), IP2 (middle) and IP5 (bottom) in the LHC, with
      the machine aperture superimposed.  The envelopes are plotted in a
      lighter color after the impact (indicated by a red star) of the
      central orbit on the aperture.  The horizontal scales have $s=0$ at
      each IP in turn
      and the closest horizontal collimators are represented as black boxes at
      $6\sigma$ to the right.  The length of the collimators is not to scale
      for the sake of readability.}\label{fig:off-mom-env-2d}
\end{figure*}

The risk of inducing quenches means that it is vital to study, quantify and possibly alleviate the impact of BFPP and EMD   in the LHC, in order to ensure safe operation uninterrupted by lengthy quench-recovery procedures.

To study the loss processes, we use a three-step simulation method, consisting of optical tracking of the ions with modified rigidity, a Monte Carlo simulation of the hadronic and electromagnetic shower created as the ions fragment following their initial impact on the vacuum envelope, and a thermal network simulation of the heat flow inside the surrounding cryo-magnet. To illustrate the method we apply it to BFPP in the \alice\  experiment at IP2 of the LHC, with brief comparisons with the other IPs in Sec.~\ref{sec:tracking}-\ref{sec:operating-cond}. In Sec.~\ref{sec:emd-quench} we treat losses from EMD and in Sec.~\ref{sec:monitoring} we describe a system of beam loss monitors (BLMs) to survey the losses. Finally, in Sec.~\ref{sec:alleviation}, we discuss  possible methods of alleviation.

\section{Optical tracking} \label{sec:tracking}

The condition for loss of nuclei modified by BFPP or EMD in a localized spot, given by Eq.~(\ref{eq:loss-condition}), depends thus on  the beam optics, the aperture and the ion species. For \pb\ ions in the LHC, the $\delta$ caused by BFPP is $\dBFPP=0.012$  (using $\Delta Z=-1$ in Eq.~(\ref{eq:delta-rigidigy})). For EMD1 ($\Delta A=-1$) and EMD2 ($\Delta A=-2$) we include also the average momentum deviation per nucleon caused by the recoil, which we estimated with \FLUKA\ to $\langle \delta_p \rangle \approx -7.4\times 10^{-5}$ for EMD1 and $\langle \delta_p \rangle \approx -1.25\times 10^{-4}$ for EMD2. This results in $\dEMDone=-0.0049$ and $\dEMDtwo=-0.0097$ from Eq.~(\ref{eq:loss-condition}).

Since the momentum acceptance of the LHC arcs is $|\delta|<0.006$~\cite{lhcdesignV1}, BFPP and EMD2 will cause localized losses in the LHC while EMD1 will not.

To illustrate the losses from BFPP and EMD, we used the program \MADX~\cite{madx} to compute the off-momentum optical functions. The computation was done for both beams using 
V6.503 of the LHC optics. Fig.~\ref{fig:off-mom-env-2d} shows the resulting beam envelopes for Beam~1 (circulating clockwise when viewed from the top) for the nominal beam and the secondary beams of ions created by BFPP or EMD at each of the experimental IPs.

Even if ion species other than \pb\ were  used, particles affected by BFPP  would still be lost, since  Eq.~(\ref{eq:delta-rigidigy}) and $|\delta|<0.006$ 
requires $Z_0 \geq 168$.  The power drops rapidly with $Z^7$ according to  Eq.~(\ref{eq:sigmabfpp}). In comparison, the corresponding condition at RHIC is $Z_0 \geq 58$  so that the measurements in Ref.~\cite{prl07} were only possible with $^{63}\mathrm{Cu}^{29+}$ beams and not the more usual $^{197}\mathrm{Au}^{79+}$.

Ions created by EMD1  are lost if $A_0\leq166$, meaning that this might be a source of localized losses during future collisions between lighter ions. However, for lighter ions, the cross section is significantly lower~\cite{pshen01}.

Since BFPP is the most dangerous process in the LHC we focus on it here and treat EMD in Sec.~\ref{sec:emd-quench}.

A \pb\ ion, in the center of the bunch and following the ideal trajectory through the center of all magnets, enters the IP with $\delta=0$.  If it acquires an extra electron through BFPP, it exits with $\delta=\dBFPP$.  Downstream of the IP it follows a dispersive trajectory $\xbfpp$, which together with the linear optics is calculated exactly with \MADX\ rather than using the linear approximation $\xbfpp\approx d\delta$.

To write the orbit of any other \pbone\ ion, we use subscript $i$ to denote that the function $A$ should be evaluated at $s=s_i$ and express derivatives with respect to $s$ as $dA/ds=A'$.  Furthermore, we use a tilde to represent chromatic optical functions for the BFPP particles (e.g. at $s=s_i$ the usual envelope function is $\beta_i$ for the nominal \pb\ beam and $\tilde{\beta}_i$ for the \pbone\ beam). Unless stated otherwise, all optical functions refer to the horizontal plane.

With $s=s_0$ at the IP, we write the position $x_1$ and angle $x_1'$ of a \pbone\ ion at some downstream position $s_1$ as
\begin{equation}
\label{eq:x-at-impact}
\begin{bmatrix}x_1 \\ x'_1 \\ \delta_p \end{bmatrix}
=\begin{bmatrix}\xbfppI \\ \xbfppI' \\ 0 \end{bmatrix}
+\begin{bmatrix}\tilde{C}_1 & \tilde{S}_1 & \tilde{d}_{1} \\ \tilde{C}'_1 & \tilde{S}_1' & \tilde{d}'_{1} \\ 0 & 0 & 1
\end{bmatrix}
\begin{bmatrix}x_0 \\ x'_0 \\ \delta_p \end{bmatrix}
\end{equation}
using a standard representation of the transfer matrix in terms of the optical functions. The trajectory of a BFPP ion is thus the superposition of the central dispersive trajectory $\xbfpp$ caused by the electron capture, a betatron oscillation around it, and a dispersive contribution from the momentum deviation within the bunch.

The functions $\tilde{C}_1$ and $\tilde{S}_1$ are given by
\begin{eqnarray}\label{eq:CS}
\tilde{C}_1&=&\sqrt{\frac{\tilde{\beta}_1} {\tilde{\beta}_0}} \left[\cos(\tilde{\mu}_1-\tilde{\mu}_0) + \tilde{\alpha}_0
\sin(\tilde{\mu}_1-\tilde{\mu}_0) \right] \nonumber\\
\tilde{S}_1&=&\sqrt{\tilde{\beta}_1 \tilde{\beta}_0} \sin{(\tilde{\mu}_1-\tilde{\mu}_0)},
\end{eqnarray}
where $\tilde{\mu}$ is the off-momentum betatron phase and $\tilde{\alpha}=-\tilde{\beta}'/2$. They are evaluated with initial conditions taken from the periodic on-momentum solution at the IP ($\tilde{\beta}_0=\beta_0$ and $\tilde{\mu}_0=\mu_0$). For the off-momentum dispersion function $\tilde{d}$, the initial condition is $\tilde{d}_0=0$.

Since the betatron amplitudes 
 and $\delta_p$ are small, and there are essentially no nonlinear elements in this part of the ring, the linear approach in Eq.~(\ref{eq:x-at-impact}) is accurate as long as the central dispersive trajectory $\xbfpp$ is calculated without approximation.

The $\delta_p$ of the BFPP particles has a Gaussian distribution with a standard deviation given in Tab.~\ref{tab:lhc-beams}. The initial $x_0$ and $x'_0$ are also normally distributed but we  must take into account that the distribution of these collision points is \emph{not} that of   the incoming bunches. Assuming zero periodic dispersion and   head-on collision between identical Gaussian bunches at an IP, the horizontal phase space distribution of each bunch is
\begin{equation}
f_\beta(x_0,x_0')=\frac{N_b \beta_0}{2\pi
\sigma_0^2}\exp{\left(-\frac{x_0^2+(\alpha_0 x_0+\beta_0 x'_0)^2}{2
\sigma_0^2}\right)}.
\end{equation}
where $N_b$ is the number of particles per bunch, $\sigma_0=\sqrt{\beta_0 \epsilon}$ and $\epsilon$ is the horizontal geometric emittance. The luminosity density $\lambda$ in phase space, which corresponds to the distribution of the BFPP ions, is then calculated by integrating over the two colliding bunches, where we denote the coordinates in the opposing bunch at the IP by $u$ and impose $u_0=x_0$ as a condition for a collision to take place:
\begin{eqnarray}\label{eq:Pbfpp}
\Pbfpp(x_0,x_0')&=&
\frac{\int f_\beta(x_0,x'_0)f_\beta(x_0,u_0') \,du_0'}
     {\int f_\beta(x_0,x_0')f_\beta(x_0,u_0')\, dx_0' \, du_0'\, dx_0}
\nonumber   \\
       &=&\frac{\beta_0}{\sqrt{2}\pi \sigma_0^2}\mathrm{e}^{-\frac{2x_0^2+(\alpha_0 x_0+\beta_0 x_0')^2}{2\sigma_0^2}}.
\end{eqnarray}
This is again a Gaussian distribution, but with a smaller standard deviation
\begin{equation}
\StandDevBfpp=\left(\int x_0^2\, \Pbfpp(x_0,x_0')\, dx_0' \,dx_0\right)^{1/2}=\frac{\sigma_0}{\sqrt{2}}.
\end{equation}
Similarly, the standard deviation of the angular distribution of the BFPP ions is
\begin{equation}
\sigma_{p,0}=\sqrt{\frac{\epsilon}{\beta_0}\frac{2+\alpha_0^2}{2}},
\end{equation}
which reduces to the familiar expression $\sqrt{\epsilon/\beta_0}$ if $\alpha_0=0$ (as holds at all IPs in the LHC). We see that the distribution of collision points in space is narrower than the bunch distribution by a factor $\sqrt{2}$, while the angular distribution is similar to that of the initial bunch. The vertical coordinates $y$ and $y'$ are treated analogously.

The size of the BFPP beam changes as it propagates through the accelerator lattice.  Using Eq.~(\ref{eq:x-at-impact}) to propagate the beam distribution in Eq.~(\ref{eq:Pbfpp}) the standard deviation of the BFPP particles at $s_1$ can be shown to be
\begin{equation}
\label{eq:BFPP-beam-size}
\StandDevBfppOne=
     \sqrt{\frac{1+\sin^2(\tilde{\mu}_1-\tilde{\mu}_0)}{2}\tilde{\beta}_1\epsilon
           +\tilde{d}_1^2(s) \langle \delta_p^2 \rangle},
\end{equation}
This can be compared to the main beam, with $\sigma_1=\sqrt{\beta_1\epsilon+D_1^2(s)\langle\delta_p^2\rangle}$, where $D_1$  is the periodic on-momentum dispersion. Differences come both from the fact that the BFPP beam envelope is oscillating with $\tilde{\mu}$, rather like a mismatched beam at injection, and  the chromatic variation of the optical functions.

We can deduce the impact point of the BFPP losses directly from Fig.~\ref{fig:off-mom-env-2d}.  Once this is known, we implement a fast tracking of single particles from the IP to the beginning of the element where losses occur using Eq.~(\ref{eq:x-at-impact}). Inside this element an analytical algorithm is used to find the impact coordinates and momenta.

\begin{figure}
  \includegraphics[width=7cm]{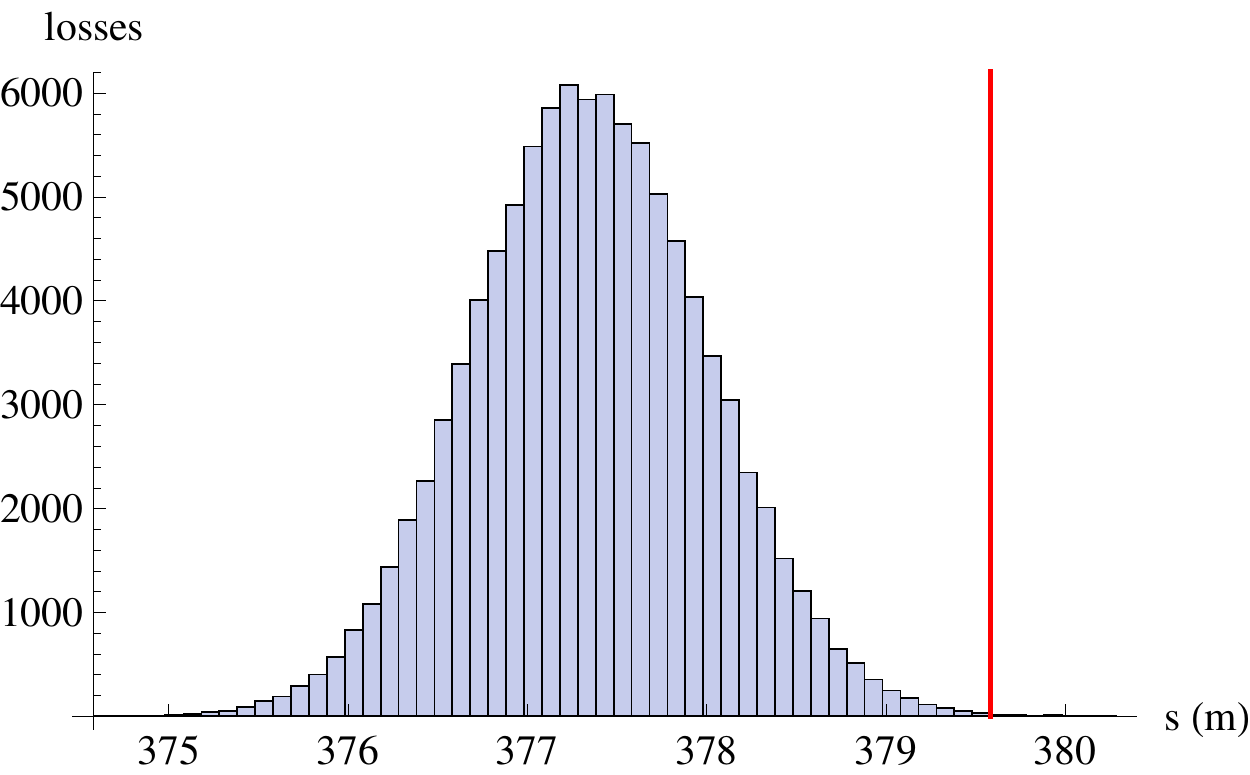}\\
  \caption{The longitudinal distribution of the impact positions of lost
  \pbone\ ions from BFPP after IP2 in the LHC, as simulated by
  tracking 10$^5$ particles. The red line indicates
  the end of a superconducting dipole magnet.}\label{fig:impacts-IP2}
\end{figure}

\begin{figure}
  \includegraphics[width=8.5cm]{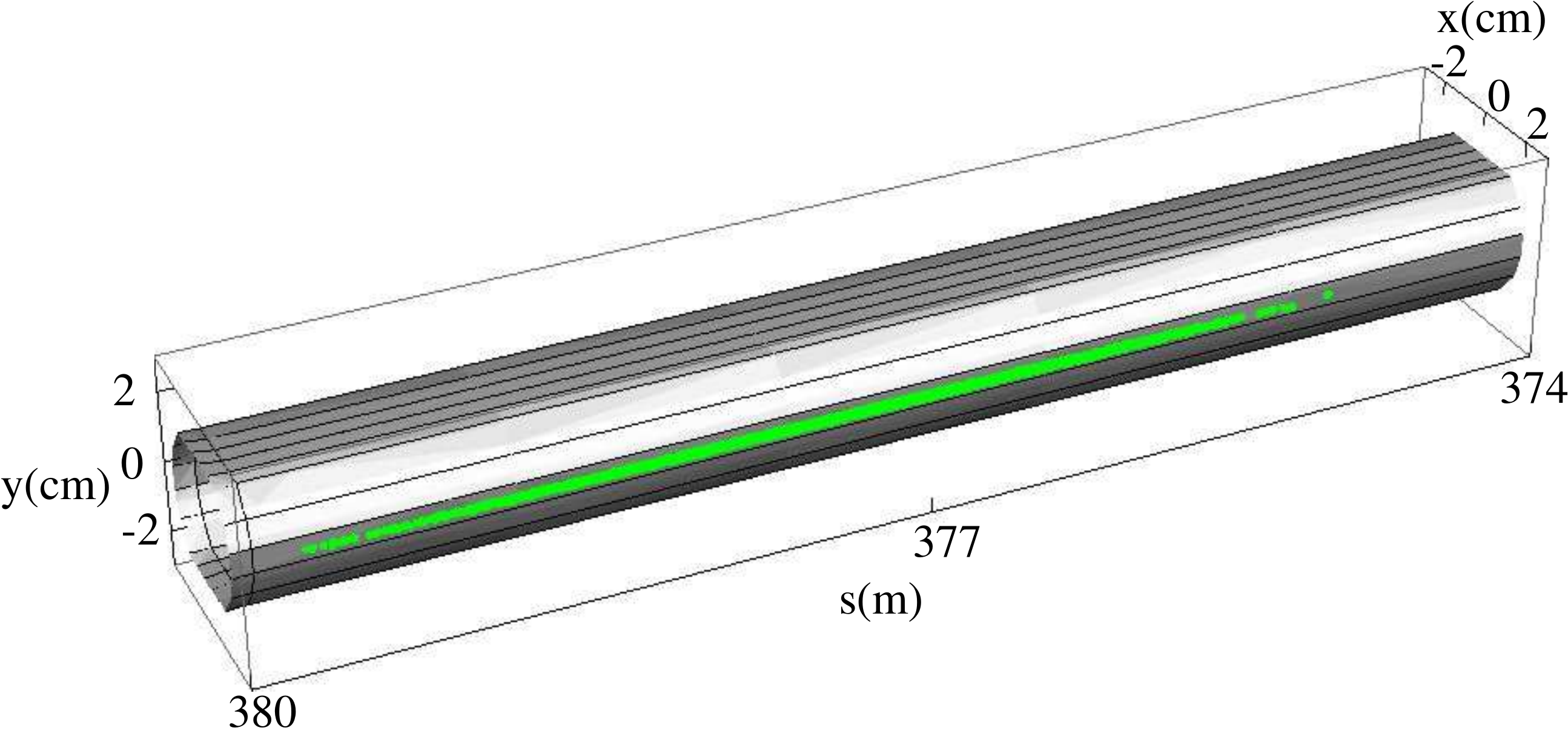}
  \caption{The distribution of the impact positions of lost
  \pbone\ ions (green dots) from BFPP after IP2 in the LHC, as
  simulated by tracking 10$^4$ particles,
  shown together with the beam screen. 
}\label{fig:pipe-Impacts}
\end{figure}

At IP2 in the LHC, the BFPP beam converted from the clockwise-circulating Beam 1 is lost near the end of a superconducting dipole magnet in the dispersion suppressor (known as ``MB'' type). Tracking gives normally distributed losses along $s$, shown in Figs.~\ref{fig:impacts-IP2} and~\ref{fig:pipe-Impacts}, with a mean of $s=377.35$~m from IP2 and a standard deviation of 65~cm. In the case of IP1 or IP5 the loss positions are    around 418~m downstream of each IP, also in superconducting dipole magnets, with projected rms spot sizes of 58~cm and 74~cm respectively. The situation around the three IPs is therefore comparable, and we focus on IP2 in the remainder of this paper, except where otherwise indicated.

\begin{figure}
  \includegraphics[width=8cm]{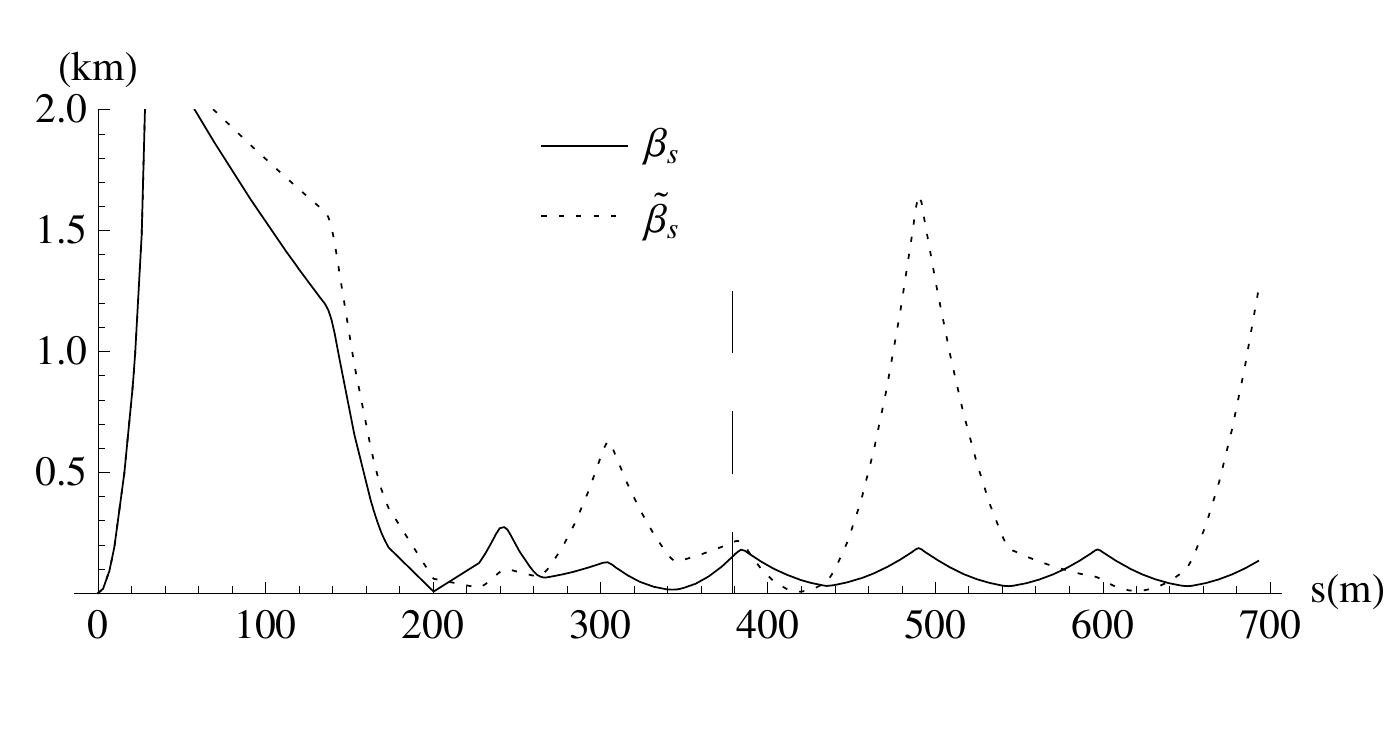}\\
  \includegraphics[width=8cm]{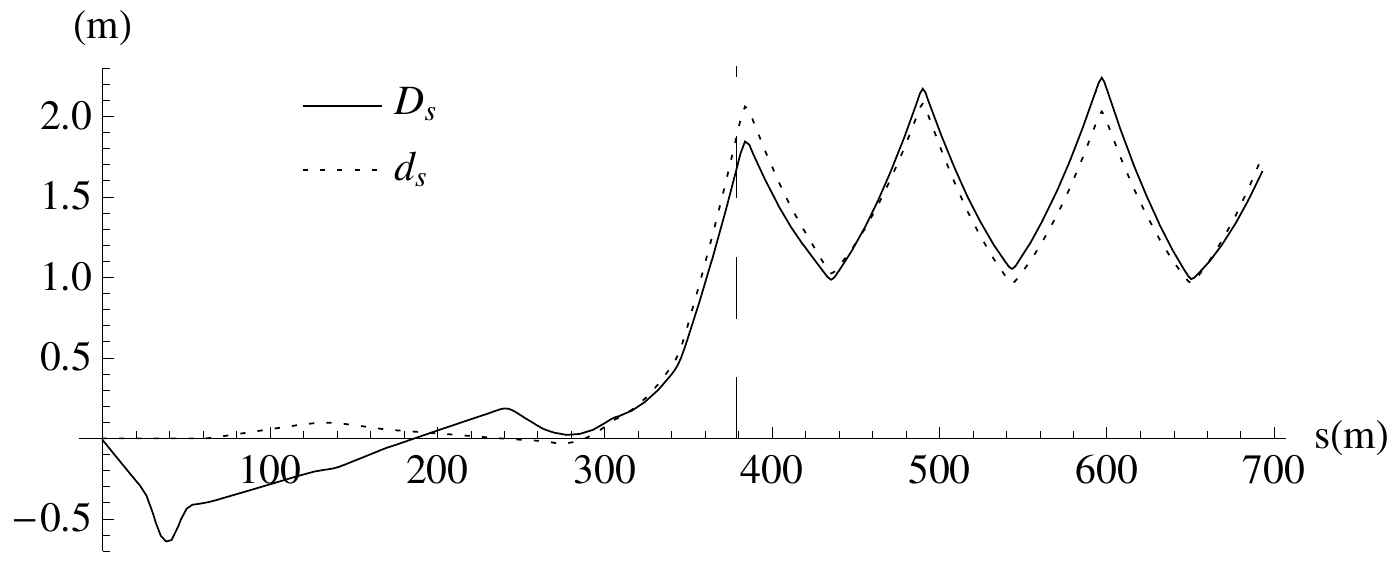}\\
  \caption{The on-momentum horizontal $\beta$-function downstream of
          IP2 compared with the off-momentum $\tilde{\beta}$, calculated as
          described in the text (up), and the periodic dispersion function $D$ shown together with the
          locally generated  dispersion $d$ (down).
          The dashed vertical line indicates the central loss position
          of the BFPP particles,
          close to the first maximum of the dispersion.
          }\label{fig:off-mom-optics}
\end{figure}

In Fig.~\ref{fig:off-mom-optics} we show the on- and off-momentum $\beta$-functions downstream of IP2.  Around the impact point we have $\beta_1\gg\tilde{\beta}_1$ meaning that the BFPP beam is much smaller than the nominal beam.  This effect concentrates the heat load in a  smaller volume of the magnet.

\section{Shower simulation} \label{sec:bfpp-shower}

\begin{figure}
  \includegraphics[width=6.2cm]{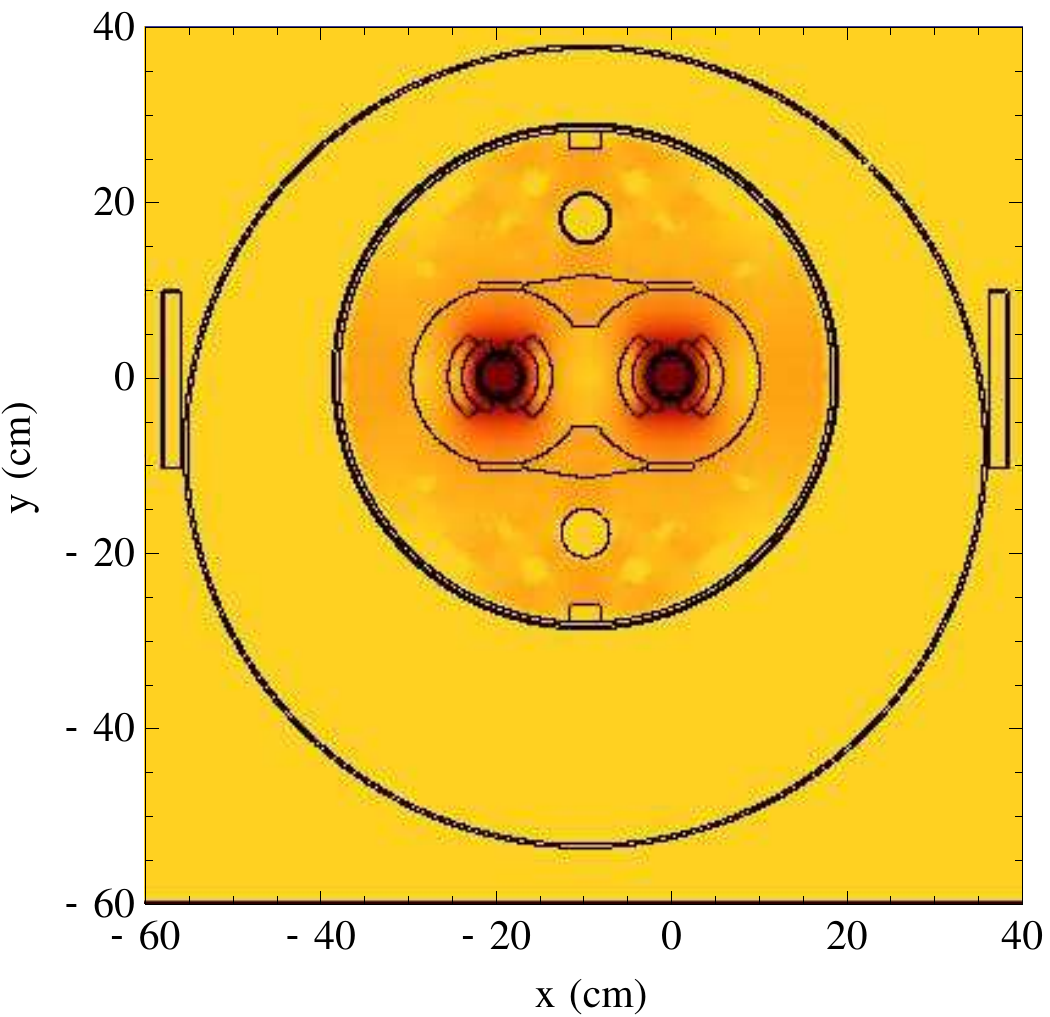}
  \includegraphics[width=1.3cm]{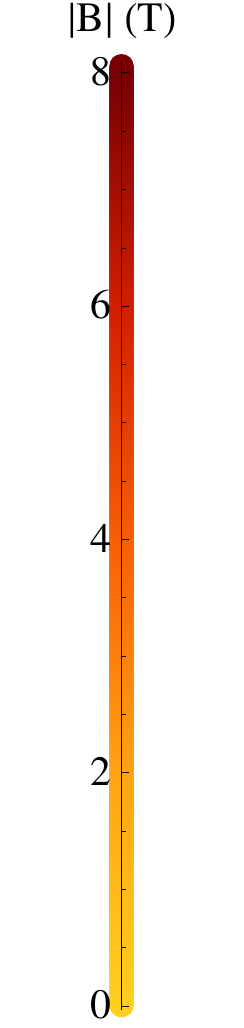}\\
  \includegraphics[width=8.5cm]{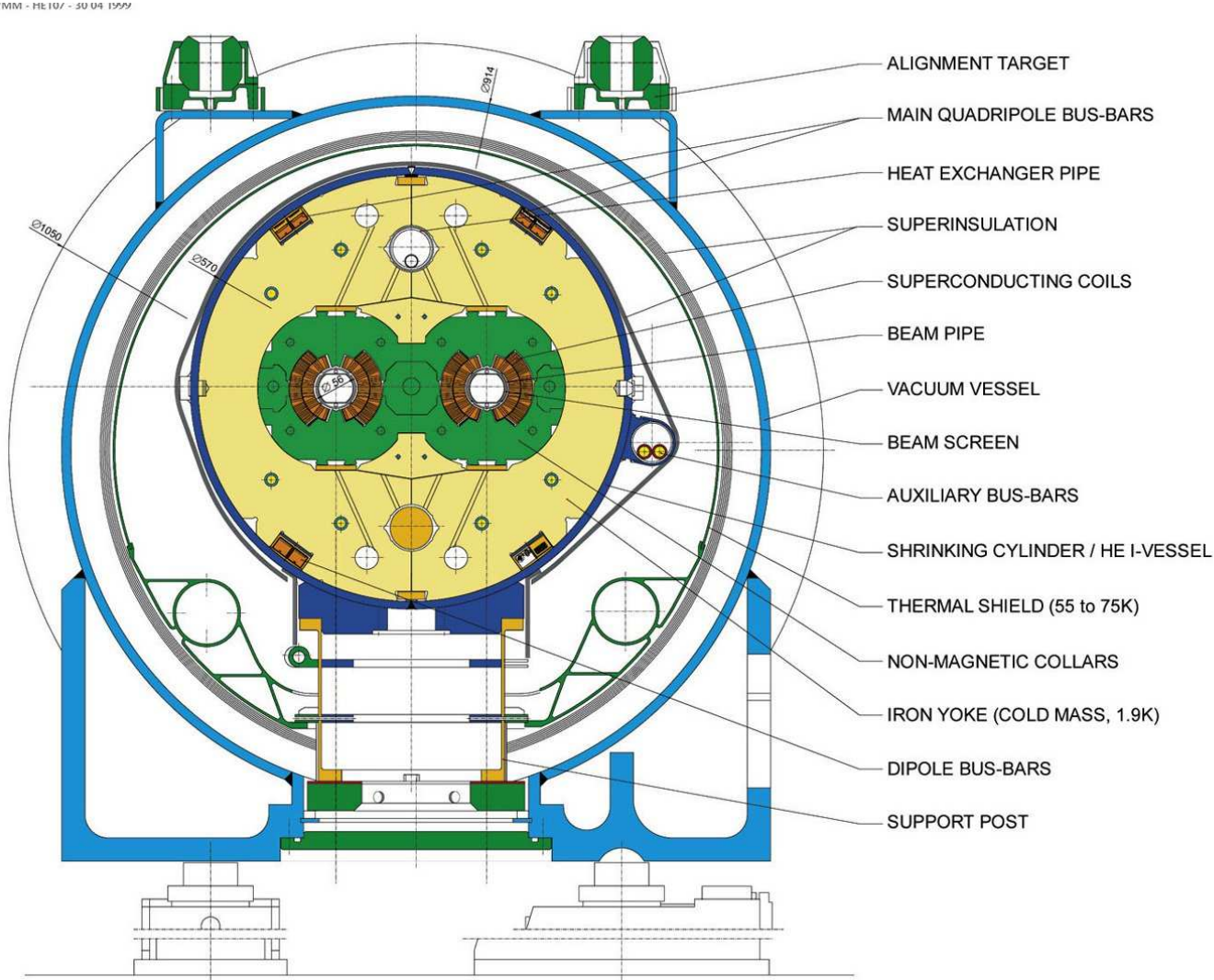}\\
  \includegraphics[width=8.5cm]{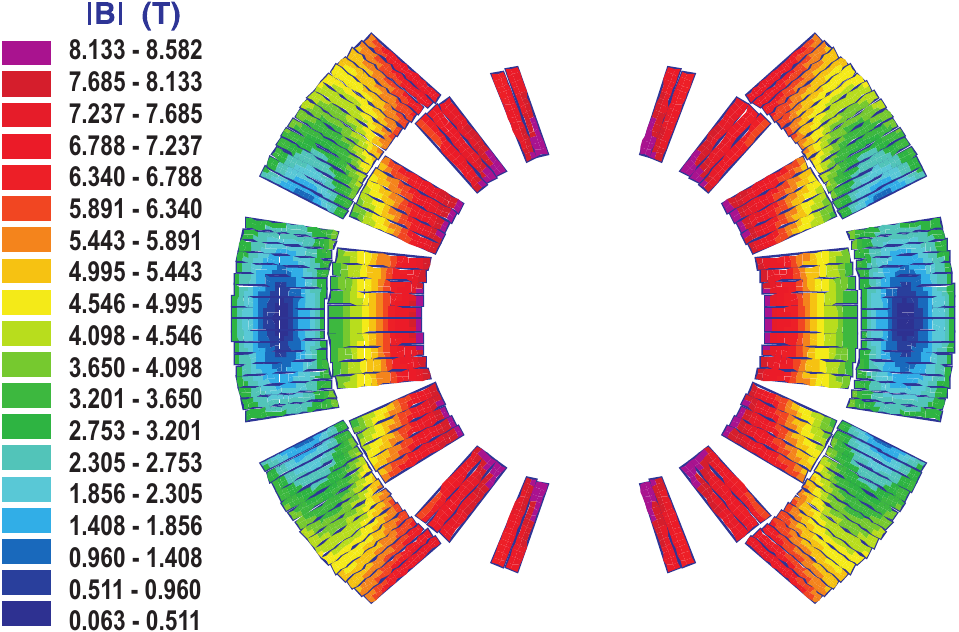}\\
  \caption{The transverse cross section of the \FLUKA\ model with the magnetic field superimposed (up),
           a technical drawing~\cite{lhcdesignV1} (down) of an LHC main dipole (middle) and the magnetic field in the superconducting cables (down). The
  rectangular boxes on the outside of the magnet in the \FLUKA\ model represent beam-loss
monitors as discussed in Sec.~\ref{sec:monitoring}.}\label{fig:lhc-MB-geo}
\end{figure}

The next step  to determine the risk of quenches from the BFPP beam losses  is to estimate the energy deposition they give rise to. Defining the origin of cylindrical coordinates $(r,\phi,z)$ at the center of the beam pipe at a magnet entrance, the heating power density $P$ is \begin{equation} \label{eq:heat-power} P(r,\phi,z)=\lum \,\sigbfpp \,W(r,\phi,z), \end{equation} where $W$ is the average, over many possible showers, of the energy deposition per unit volume in the superconductor or other material per lost \pbone\ ion.

The quantity $W$, which has the unit $\mathrm{J}/(\mathrm{cm}^3\,\mathrm{ion})$, depends on the distribution in space and momentum of the ions lost on the beam pipe, and the cross sections for the many interactions that occur as the shower develops in the material structure of the magnet. We estimate  $W$ through simulations  with \FLUKA, where a 3D model of a main dipole has been implemented~\cite{magistris06} as shown in Fig.~\ref{fig:lhc-MB-geo}.

The lost particles first hit the beam screen, which is a racetrack-shaped chamber intended to protect the magnet from synchrotron radiation. Outside is the circular cold bore (labelled ``beam pipe'' in Fig.~\ref{fig:lhc-MB-geo}), which is surrounded by the superconducting coil, consisting of an inner and an outer winding.  The collar is a part of the coil support structure and is inserted in the iron yoke.  All these parts are heated by the hadronic and electromagnetic showers.

The \FLUKA\ model of the magnet contains some simplifications:
\begin{itemize}

\item The matrix of superconducting NbTi filaments and copper inside the coil was modeled as a homogeneous body with weight fractions of the materials corresponding to the real coil.

\item The curvature of the magnet was neglected (although it was included in the tracking up to the impact on the beam screen). The cold mass is 15.2~m long and has a sagitta of 9.18~mm~\cite{lhcdesignV1}. Simplified FLUKA simulations, representing the magnet as a
\clearpage 
solid block (either rectangular or with a 9.18~mm sagitta), shows that the introduced error is around 3--4\%, which is negligible compared to the overall simulation error.

\item The magnetic field was included inside the yoke but omitted outside, where it cannot affect the results of the magnet quench evaluation.

\item Parts of the magnet far away from the coil are not necessary
to include for the purpose of estimating the energy deposition in the
coils, however, they are needed for later simulations of the beam loss
monitors (BLMs) described in Sec.~\ref{sec:monitoring}.

\end{itemize}

The coordinates and momenta of the particles impinging on the inside of the beam screen, generated by the tracking described in Sec.~\ref{sec:tracking}, were fed as
initial conditions to the \FLUKA\ simulation. Some $10^4$~particles were sufficient to keep the statistical error below 2\% when the energy deposition in each cell of
several spatial cylindrical grids was recorded.

\begin{figure}
  \includegraphics[width=7.5cm]{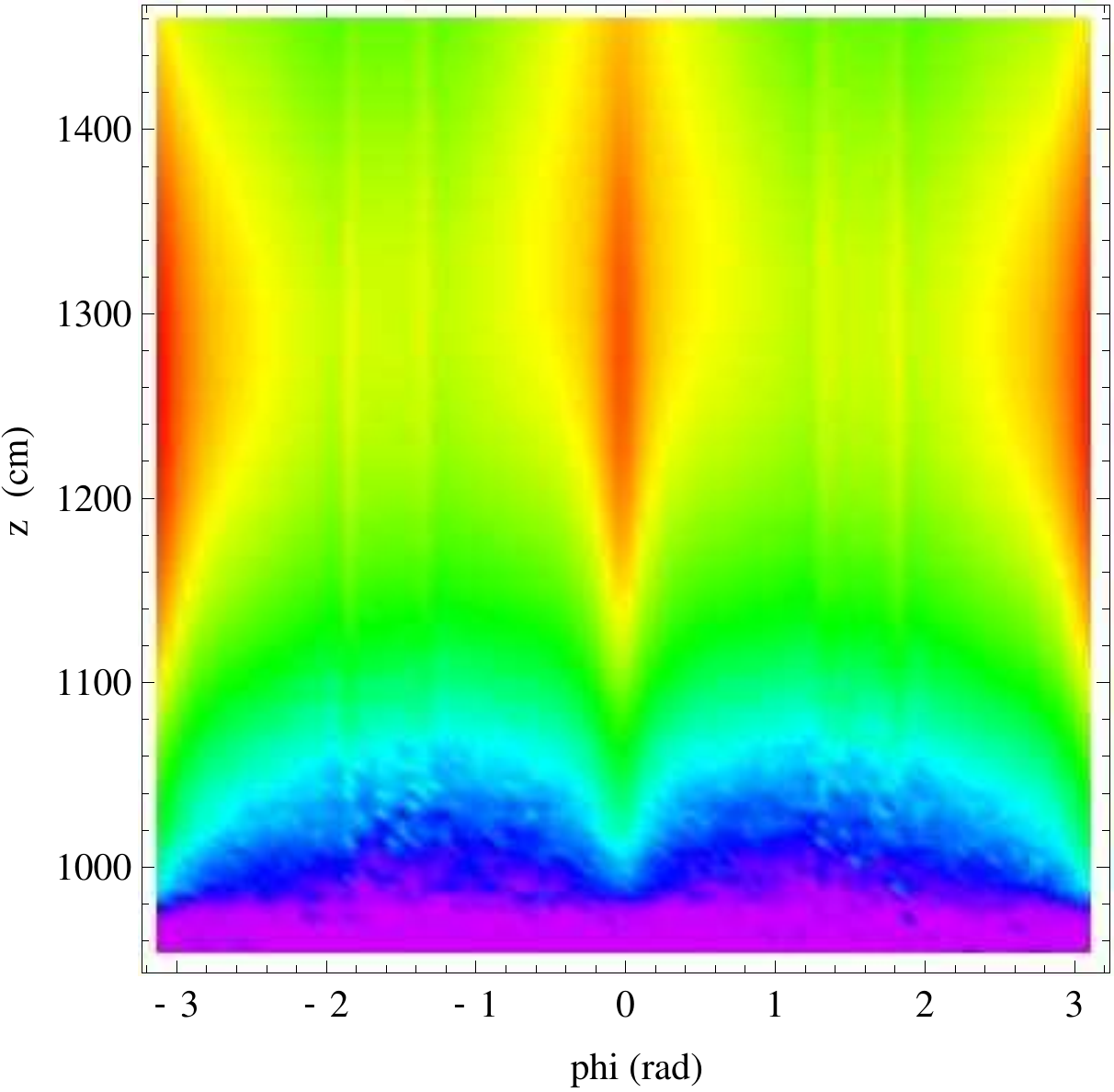}\\
\begin{center}
  \includegraphics[width=6.5cm]{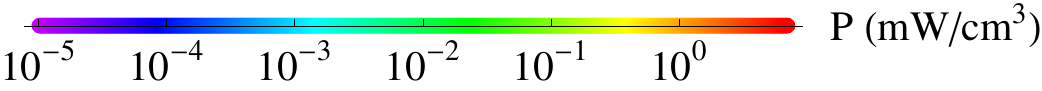}\\
\end{center}
  \caption{
The heating power from beam losses caused by BFPP in the inner layer of
the coil of an LHC main dipole as simulated with \FLUKA.
The power density was
averaged over the width of the cable and is shown as a function of
azimuthal angle $\phi$ and longitudinal coordinate $z$, with $z=0$ in
the beginning of the magnet. The beam loss is centered around $z=1206$~cm and $\phi\approx -3.11$~rad. }\label{fig:shower-inner-coil}
\end{figure}

\begin{figure}
  \includegraphics[width=8cm]{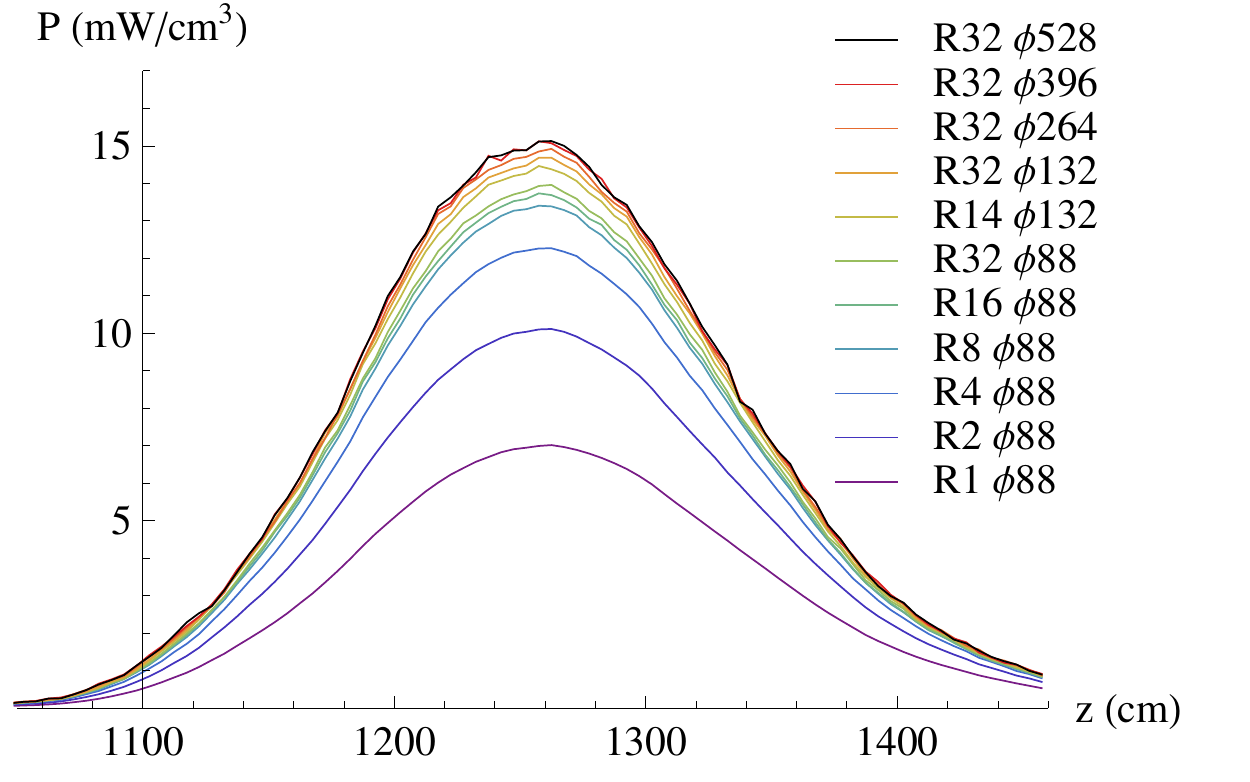}\\
  \caption{The linear power density along $z$ caused by BFPP in the inner
layer of the coil of the dipole magnet for varying radial and azimuthal
binnings, where $Ri\;\phi j$ stands for $i$ radial bins and $j$
azimuthal      bins.
In all cases, we used a longitudinal cell size of
$\Delta z=5\,\mathrm{cm}$.
The hottest
bin was selected for each
mesh.}\label{fig:edep-wire}
\end{figure}

Fig.~\ref{fig:shower-inner-coil} shows the  power density in the inner layer of the superconducting coil and Fig.~\ref{fig:edep-wire}  presents the energy deposition along the length of  the cable at the hottest azimuth in the radial bin in the coil closest to the impact for different meshes. In both figures, energy deposition is converted to power density at the design luminosity of Table~\ref{tab:lhc-beams} with Eq.~(\ref{eq:heat-power}). The maximum power density  in the coil is $P\approx 15.5\,\mathrm{mW/cm^3}$ using LHC optics V6.503, and does not change if the cell sizes $r\Delta r\,\Delta z\,\Delta \phi$ are reduced.

We estimate that the \FLUKA\ simulation has a systematic error margin of a  factor~2--3, taking into account the uncertainties from the surface effects in the grazing incidence, the transverse momentum distribution, and the showering and cross sections at LHC energy. This error is consistent with Refs.~\cite{prl07,prstabSPS09}.

\section{Thermal network simulation} \label{sec:bfpp-network}

The heat load distribution in the coil can be used to estimate the resulting temperature profile. Since the flux of lost \pbone\ ions changes only on the scale of tens of minutes, a steady state situation is considered, with a continuous deposition and evacuation of heat from the coil. We discuss fluctuations of the heat load in the end of this section.

\begin{figure}
  \includegraphics[width=7cm]{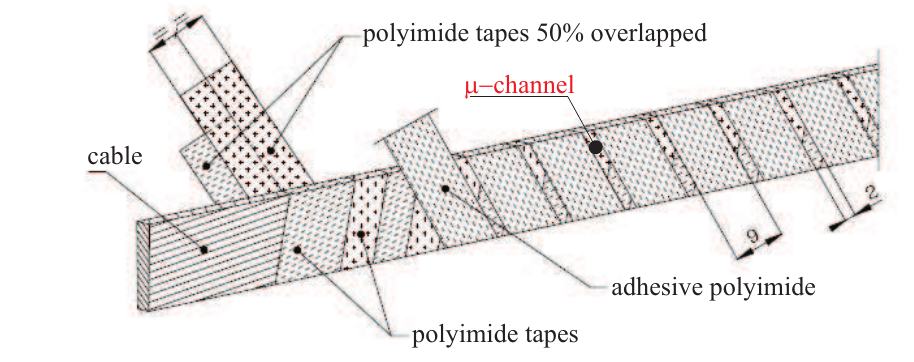}\\
  \includegraphics[width=7cm]{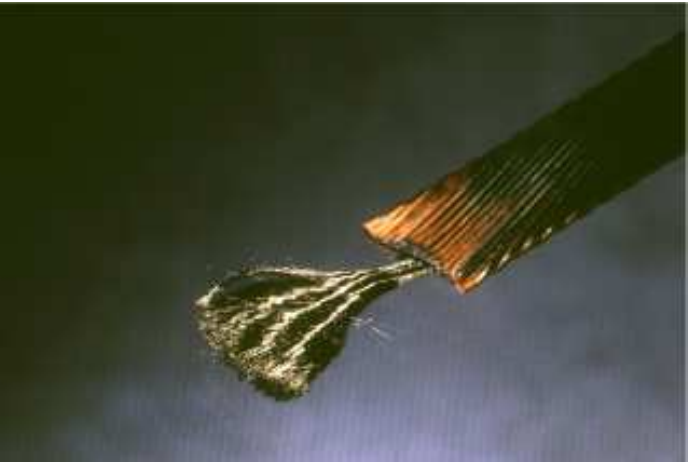}\\
             \caption{Close-up of a single cable where the current carrying strands,
             and the filaments composed of NbTi, are visible.}
             \label{fig:sc-cable}
\end{figure}

\begin{figure}
  \includegraphics[trim = 10mm 50mm 110mm 5mm, clip, width=6.5cm]{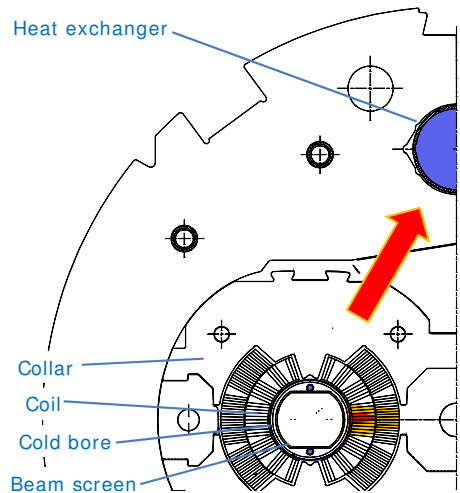}
  \caption{Cross section of a part of an MB magnet where the red arrow schematically indicates the heat flow from the coil to the heat exchanger.}
  \label{fig:LHC-MB-cs}
\end{figure}

To better understand the heat flow, we describe briefly the coil geometry. The Rutherford-type cables in the coil (see Fig.~\ref{fig:sc-cable}) consist of strands (28 and 36 in the inner and outer windings respectively), made of NbTi filaments, which are embedded in a copper matrix. Helium inside the cables occupies the space between the strands~\cite{bocian03} but serves mainly to increase the heat capacity for protection against transient losses~\cite{granieri01}. The cables are wrapped with three layers of polyimide electric insulation and a \mm{0.5}-thick ground electric insulation is placed between the coil and the collar, which has been identified from measurements as a heat reservoir~\cite{bocian04}.  The collar is built from \mm{3}-thick austenitic steel plates with a  \mm{0.1} gap between them filled with helium in direct contact with the heat exchanger through the helium in the iron yoke. The heat flow in steady state, schematically shown in Fig.~\ref{fig:LHC-MB-cs}, is mainly limited by the heat conduction of the electric insulation of the cables and the size of the helium channels in the magnet.

There are several estimates of the quench limit, that is the smallest power density that could cause a quench. In Refs.~\cite{report44,lhcdesignV1} it is estimated as $5  \,\mathrm{mW/cm^3}$ and $4.5\,\mathrm{mW/cm^3}$. The maximum simulated BFPP power is more than a factor~3 higher. These estimates are based on experiments~\cite{burnod94,meuris99} where a homogeneous heat deposition in a cable sample is assumed. In the case of BFPP however, the energy deposition is non-uniform  and  distributed over the entire MB coil.

Refs.~\cite{bocian01, bocian02} describe a more detailed method of heat flow modeling in the coil, based on a thermal equivalent of an electrical network. This \emph{network model} can be used to simulate the steady state heat flow caused by a given input heat source, e.g. a beam loss. It can then be inferred from the temperature map whether the magnet quenches or not. Since the  network model  simulations in Refs.~\cite{bocian01, bocian02} show that the quench limit is highly dependent on the heat load distribution and the coil structure,   a new  network model  simulation, directly using the heat deposition map simulated with \FLUKA, should be the most accurate assessment of the risk of quenches due to the BFPP beam losses.

The network model takes as input a mesh of thermal elements, which was created from technical drawings of the magnet. It describes accurately the radial structure of the magnet from cold bore to collar, i.e., the cold bore, cold bore insulation, the helium channel around the cold bore and the coil, with the strands as the smallest thermal unit. It implements the thermal paths of the heat flowing from the cables through the insulation to the collar (Fig.~\ref{fig:LHC-MB-cs}). Furthermore, superfluid, normal fluid and gaseous helium phases are taken into account, as well as both heat conduction and convection, and nucleate boiling of normal fluid helium~\cite{bocian02}.

\begin{figure}
  \includegraphics[width=8cm]{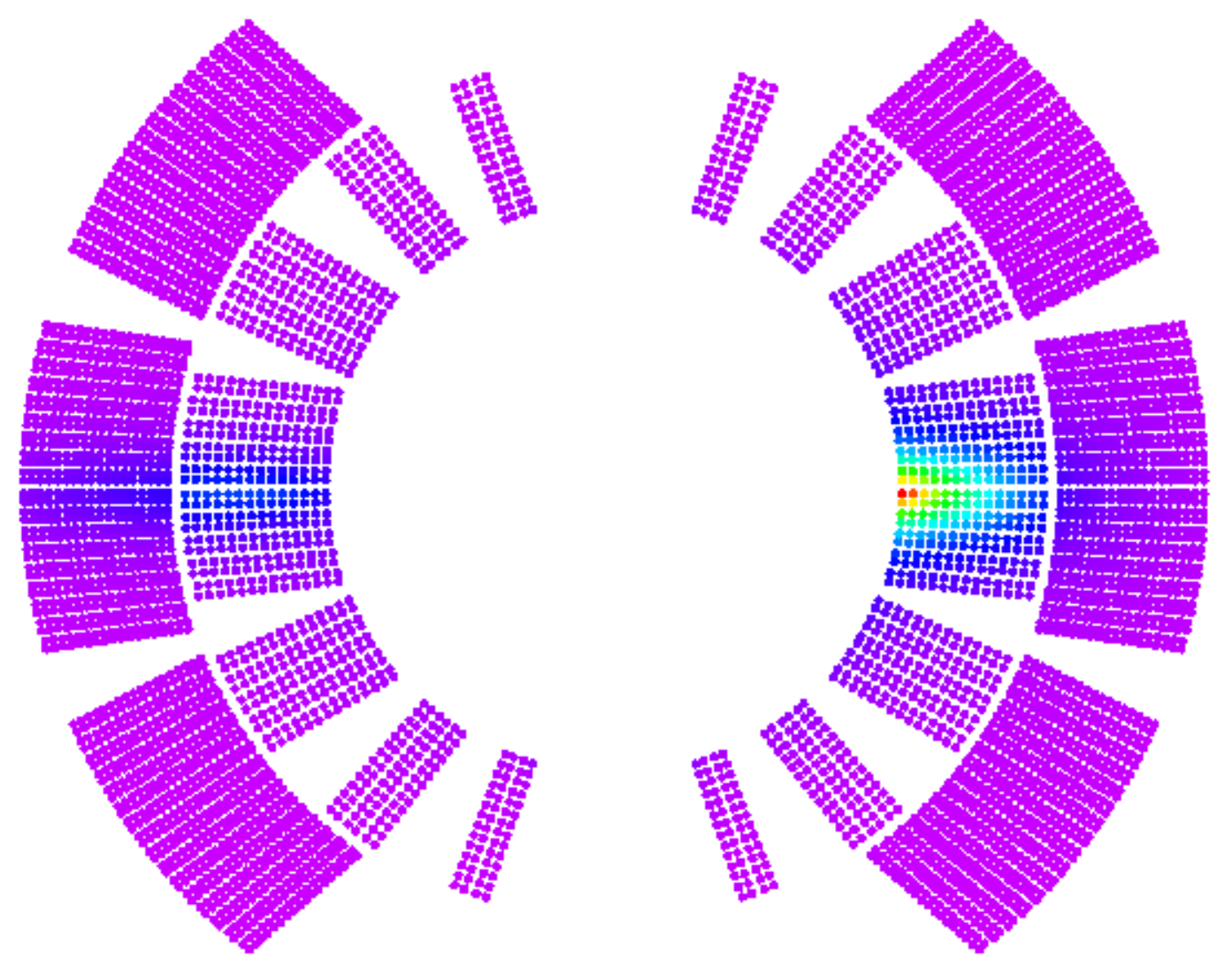}\\
  \includegraphics[width=7cm]{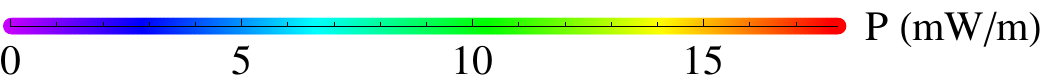}\\
\caption{The power deposition in each strand of the superconducting coil
in the 5~cm-longitudinal cell where it assumes its maximum value,
as interpolated from the result of the \FLUKA\ simulation.
}\label{fig:edep-strands}
\end{figure}

Other required simulation input includes the heat conductivity of the materials in the coil (calculated with the Cryodata software~\cite{cdata}), the temperature margin distribution in the coil, computed with the \textsc{roxie} program~\cite{russbook}, and the power deposition in each strand. Since they are not arranged in a regular polar mesh, the binning used in the \FLUKA\ simulation cannot be made to correspond exactly to the strand positions. The power in the strands is therefore obtained through a two-dimensional linear interpolation of the \FLUKA\ output (shown in Fig.~\ref{fig:edep-strands}) using the cell dimensions
$\Delta z= 5\,\mathrm{cm}$,
$\Delta \phi= 2\pi/132\,\mathrm{rad}$,
and 
$\Delta r\approx 1.1\,\mathrm{mm}$ (corresponding to 14 radial bins).

\begin{figure}
  \includegraphics[width=7.5cm]{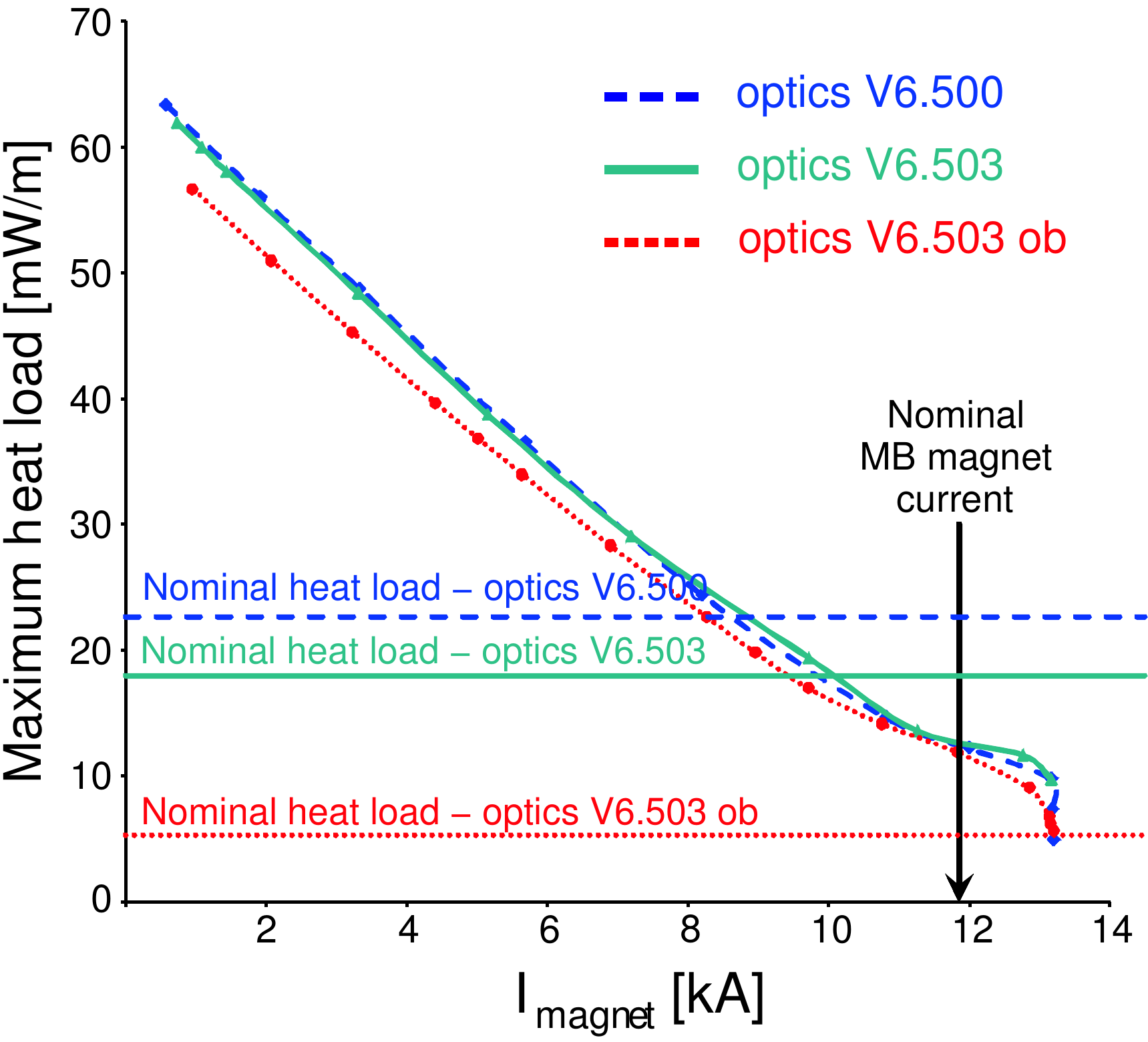}\\
\caption{Maximum allowed heat load in a single strand versus current
of the MB dipole magnet for three different spatial heat load distributions caused by BFPP
in different versions of the LHC optics: V6.500, V6.503 (current) and V6.503~ob (with an orbit bump
as discussed in Sec.~\ref{sec:alleviation}).
The horizontal lines show the heat loads expected at design luminosity as calculated by FLUKA.
The required current for 2.76~TeV/nucleon \pb operation is indicated by a vertical black arrow. At the arrow, the
tolerated power is smaller than the expected for V6.500 and V6.503,
meaning that quenches can be expected during nominal operation.
}\label{fig:iquench}
\end{figure}

In the course of the  network model simulation, the input distribution in the magnet is varied by scaling each value of the heat load map in Fig.~\ref{fig:edep-strands} up or down by a global scaling factor in order to determine the quench limit. The result of this simulation is presented in Fig.~\ref{fig:iquench}, which shows the maximum allowed input power load in a single strand before a quench occurs as a function of the magnet current for three different power load distributions caused by BFPP losses in different LHC optics versions. (This is somewhat artificial since changing the magnet current would normally amount to a change in beam energy and loss distribution but the abstraction illuminates the physics of the magnet, considered as a ``target''.) Fig.~\ref{fig:iquench} shows possible working points of the magnet for a given heat source distribution, and the margin to quench at nominal beam energy can be read out directly. Here we discuss LHC optics V6.503 and leave the others for later sections. The expected heat loads are indicated by a horizontal lines.

The function shown in Fig.~\ref{fig:iquench} drops steeply for large currents, when the current itself is the limiting factor for quenching. In this regime, the energy is efficiently transported away by the superfluid helium. This is not the case for smaller currents since, when higher temperatures are tolerated, the helium loses its superfluidity. When not all of the energy can be transported away, the heating becomes the limiting factor for quenching and the function is approximately linear.

As can be seen from Fig.~\ref{fig:iquench}, quenches are likely to occur at nominal operation and the luminosity for case V6.503 has to be decreased by  30\% to go below the quench limit. Since the loss conditions are very similar at IP1 and IP5, quenches can also be expected there at comparable levels.

When averaging the power from BFPP losses over the width of the cable, the simulated quench limits for these loss distributions are 5--6~mW/cm$^3$ depending on optics version. This is in good numerical agreement with Refs.~\cite{report44,lhcdesignV1}, although we have assumed a higher temperature margin and lower magnetic field. Therefore, for the same temperature margin, our modeling of the magnet is more pessimistic for the BFPP heat distribution.

All presented simulations assume steady state, where the heat load is constant in time. However, the real distribution of the losses in time shows statistical fluctuations. The number of BFPP particles created in a single bunch crossing can be approximated by a Poisson distribution. At nominal luminosity the expectation value is $\lum \sigbfpp/(k_\mathrm{b} f_\mathrm{rev})=0.043$ events per crossing, where $\lum$ and $k_b$ are given in Tab.~\ref{tab:lhc-beams}, $\sigbfpp$ by Eq.~(\ref{eq:sigBFPP-LHC}), and $f_\mathrm{rev}=11.1\,\mathrm{kHz}$ is the revolution frequency. The sum of BFPP particles created in several crossings has also a Poisson distribution, which can be approximated by a normal distribution for a large number of particles.

To estimate if statistical fluctuations could cause quenches, we assume operation at 95\% of the quench level (which might be overly optimistic) and calculate the probability of an increase in the number of BFPP particles large enough to induce a quench. The network model simulations show that the temperature margin at this average heat load is 0.3~K. We consider three timescales as in Ref.~\cite{granieri01}:
\begin{itemize}
 \item $t\lesssim10\,\mu \mathrm{s}$: During this time, corresponding to 65~bunch crossings, 1.9~BFPP events are expected on average, and the heat deposited in the superconducting cable is not yet transferred to the helium inside the cable. The capability of withstanding a quench is thus given by the cable enthalpy reserve corresponding to a 0.3~K increase, which is calculated to 0.48~mJ/cm$^3$. If the BFPP particles are pessimistically assumed to be lost at the same $s$-value, this corresponds to an increase in the number of BFPP events during a 10~$\mu$s interval by more than a factor~1000. The probability of this occurring during 1~month of \pb\ operation at full luminosity (which is overly optimistic) is practically zero.
\item $10\,\mu \mathrm{s}\lesssim t\lesssim0.1\,\mathrm{s}$: The heat is transferred to the helium inside the cable but not out through the insulation. The enthalpy reserve is 11.7~mJ/cm$^3$, corresponding to an increase in the number of BFPP events by a factor~8 during 0.1~s. The probability of this occurring during 1~month of \pb operation is again vanishingly small.
\item $t\gtrsim0.1\,\mathrm{s}$: Heat is transferred out through the cable insulation to the heat exchanger and we consider the steady state quench limit given by the network model. During 0.1~s there is on average 19068 BFPP events, while 20071 are required for a quench. The expected number of such fluctuations during one month is $4.8\times10^{-6}$.
\end{itemize}
The very small probability of a statistical fluctuation causing a quench justifies our steady state approach.

A successful benchmark of the network model~\cite{bocian02} used a special heating device inserted into a cold magnet to produce a known heat load. Measurement and simulation agreed within 30\% for the current where a quench occurs in the relevant range of heating the power. We take this a guide to the uncertainty of the network model simulation.

\section{Sensitivity analysis} \label{sec:sensitivity}

In reality, the impact points of BFPP ions vary as a function of imperfections such as magnet misalignments, field errors, and the imperfect aperture. Taking into account the real geometric aperture interpolated from measurements~\cite{missiaen08,missiaen09}, the longitudinal loss distribution changes slightly, but \FLUKA\ simulations show that the increase of peak heat load in the coils is negligible.

We also  studied the variations of the loss pattern caused by optical imperfections.  Imperfect corrections resulting in a residual closed orbit caused by measured magnet misalignments~\cite{missiaen08,missiaen09} were simulated and BFPP ions tracked in the resulting optics. The average longitudinal position of the impact point near IP2 varied by up to 2~m. A  negative horizontal closed orbit, $x_c$,   in  the BFPP impact zone causes BFPP ions to be lost further downstream and losses may appear in the corrector magnet and beam position monitor (BPM) attached to the next main quadrupole.  Such losses could make the BPM unusable. Unrealistic orbits $x_c \lesssim -2\,\mathrm{mm}$ at the impact point cause some BFPP ions to miss the aperture completely at the first maximum of the dispersion function and continue instead to the second (see Fig.~\ref{fig:off-mom-env-2d}), where they hit another main dipole magnet. This is discussed further in Sec.~\ref{sec:alleviation}.

To illustrate the sensitivity of the whole simulation chain with respect to optical distortions, we compared with V6.500, an earlier version of the LHC optics, with a 7\% smaller $\beta$ at the impact point after IP2. However, in the off-momentum optics, $\tilde{\beta}$ is a factor~2 smaller, reducing the longitudinal spot size to $\StandDevBfppOne\approx 49\,\mathrm{cm}$. The resulting maximum allowed power as a function of current is shown in Fig.~\ref{fig:iquench}. This is very similar to V6.503, but the expected heating during nominal operation is about 25\% higher due to the smaller spot size.

The error sources in all simulation steps 
give rise to a large uncertainty of approximately a factor~3.9 on the final quench limit in the worst case. The real error is however expected to be much smaller than that. Our practical conclusion  is that the heat load from BFPP beam losses are likely to be above the quench limit and therefore limit the luminosity of Pb-Pb collisions in the LHC.

\section{Comparison of operating configurations}
\label{sec:operating-cond}

\begin{figure}
  \includegraphics[width=4.25cm,angle=90]{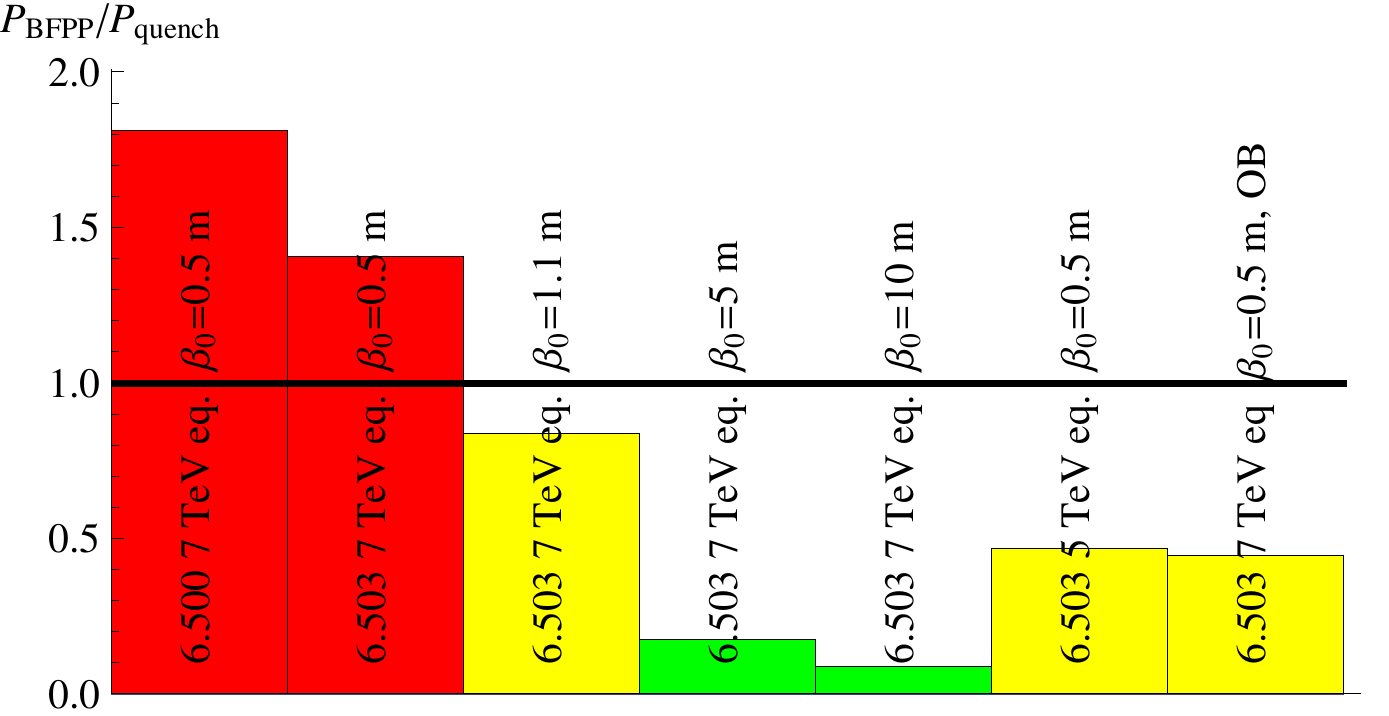}\\
  \includegraphics[width=4.25cm,angle=90]{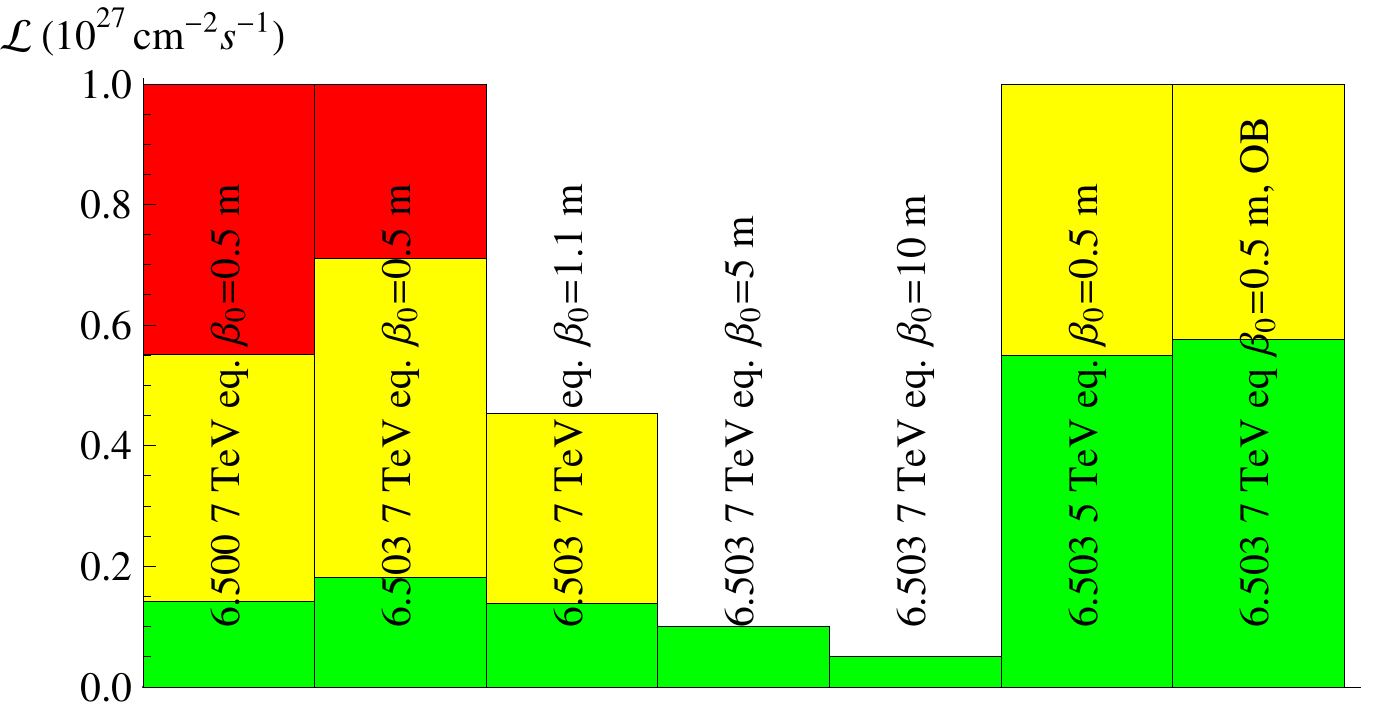}\\
\caption{The expected heating power from losses caused by BFPP at IP2 normalized by the heating power that causes a quench (top) and expected quench behavior at different luminosities (bottom) for various energies (labelled with the proton-equivalent energy), versions and configurations of the LHC optics. The colors are a semi-quantitative indication  of how dangerous the losses are expected to be:  Red bars mean that the simulated heat load is above the quench limit (note that  operation may still be possible due to simulation errors), yellow that the heat load is below the quench limit but quenches cannot be excluded due to simulation uncertainties, while green bars can be considered as safe. The height of the bars indicate design luminosity. The assumed value in the 5~TeV configuration might be too optimistic for reasons of aperture, in particular for the final focusing triplets.}
\label{fig:diff-optics-quench}
\end{figure}

As the commissioning and operation of the LHC progresses, Pb-Pb collisions will occur in a variety of configurations~\cite{lhcdesignV1,jowett08} with varying beam energy, intensity, and other parameters such as the optical function $\beta^*$ at the interaction point (note that $\beta^*=\beta_0$ defined in Eq.~(\ref{eq:loss-condition})). First ion collisions will likely take place at an energy of 1.97~TeV/nucleon (the same magnetic rigidity as a 5~TeV proton beam) rather than the nominal energy of Table~\ref{tab:lhc-beams} where magnetic fields and excitation currents are lower, meaning that more heat can be absorbed before a quench takes place (see Fig.~\ref{fig:iquench}). On top of that,   both  the luminosity and the energy deposited per lost BFPP ion  will be lower.

We repeated the tracking and \FLUKA\ simulation for this case, assuming $\beta_0=0.5\,\mathrm{m}$ at IP2 (which is over-optimistic for reasons of aperture but a useful comparison). The resulting distribution of the heat deposition is very similar to the nominal case apart from a global scaling factor of 0.63. Thus there is no need to redo the thermal network simulation---the maximum tolerated heat load can be read out directly from the curve 6.503 in Fig.~\ref{fig:iquench}, at the appropriate  current of 8.46~kA. The error introduced by omitting the last simulation step is less than 10\% if we use as a benchmark V6.500, for which the full simulation chain was carried out.  This gives an expected heat load with a 1.97~TeV/nucleon \pb beam at 46\% of the quench level, which means that quenches are unlikely to take place but still within the simulation uncertainty.

We have performed similar tracking and \FLUKA\ simulations for various values of $\beta_0$ that may be used for collisions.  Increasing $\beta_0$ reduces the luminosity ($\lum\propto \beta_0^{-1}$) and the event rate for BFPP (see Eq.~(\ref{eq:heat-power})). Furthermore, in each optics version $\tilde{\beta}$ and $\tilde{d}$ at the impact point after IP2 are slightly different, thus causing variations in the loss distributions. The heating in the magnet scales therefore only approximately as $P(x,y,z)\propto\beta_0^{-1}$.

The results of the quench performance for BFPP at IP2 for all optics versions are summarized in Fig.~\ref{fig:diff-optics-quench}. The bottom part shows expected luminosity limitations, where we have assumed that the spatial heat load distribution in a certain distribution stays constant when the luminosity varies. This is strictly true only when the luminosity is changed through a decrease in bunch population or number of bunches.

In Fig.~\ref{fig:diff-optics-quench} we also show the result for the nominal optics but with an orbit bump introduced to reduce the maximum energy deposition. This is discussed in detail in Sec.~\ref{sec:alleviation}.

No heavy ion collisions are planned to take place at injection energy.

\section{Losses from EMD in the LHC}\label{sec:emd-quench}

Since the ions affected by EMD1 stay within the momentum acceptance of the LHC ring, they are intercepted by the momentum collimation system (see Fig.~\ref{fig:off-mom-env-2d}). However, EMD2 \emph{does} create particles outside the acceptance which are then lost in localized  spots, as shown in Fig.~\ref{fig:off-mom-env-2d}. Since the cross section is almost a factor~10 lower than for BFPP (see Eq.~(\ref{eq:sigBFPP-LHC}) and Eq.~(\ref{eq:sigEMD2-LHC})), one might expect that these losses should not cause any quenches.  However, this is not the case  because of some particular features of the machine.

\begin{figure}
  \includegraphics[width=8cm]{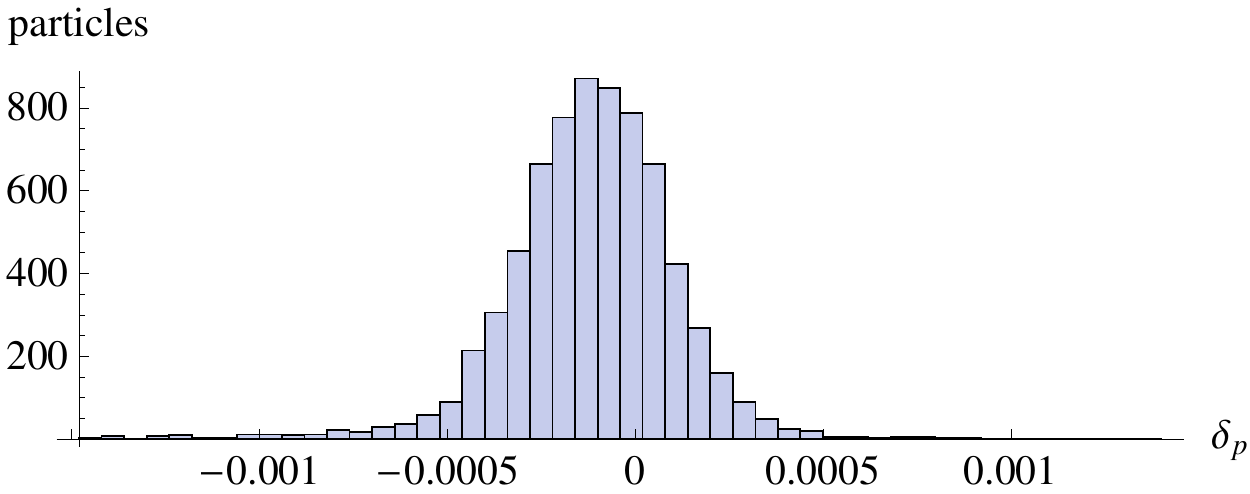}\\
  \caption{The distribution of $\delta_p$ caused by EMD2 as simulated by \FLUKA.
The distribution has $\langle \delta_p \rangle \approx -1.25\times10^{-4}$ and
$\sqrt{\langle (\delta_p - \langle \delta_p \rangle )^2 \rangle} \approx 2.6\times10^{-4}$}
\label{fig:EMD2-delta-spectrum}
\end{figure}

For EMD2 ions, we consider $\delta_p$ as the sum of two random variables: The natural momentum deviation in the bunch with standard deviation $\sigma\approx1.1\times10^{-4}$~\cite{lhcdesignV1} and the recoil caused by EMD, which was simulated with \FLUKA\ (see Fig.~\ref{fig:EMD2-delta-spectrum}). The resulting distribution is approximately Gaussian with mean $\mu=-1.25\times10^{-4}$ and standard deviation $\sigma=2.9\times10^{-4}$.

\begin{figure}
  \includegraphics[width=8cm]{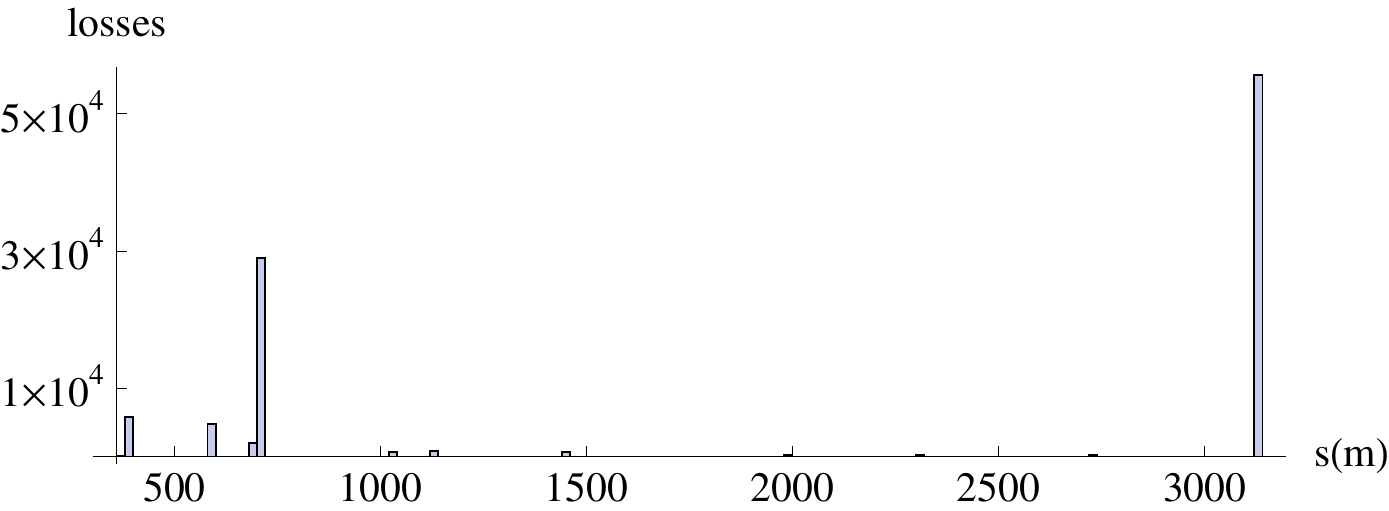}\\
  \includegraphics[width=4cm]{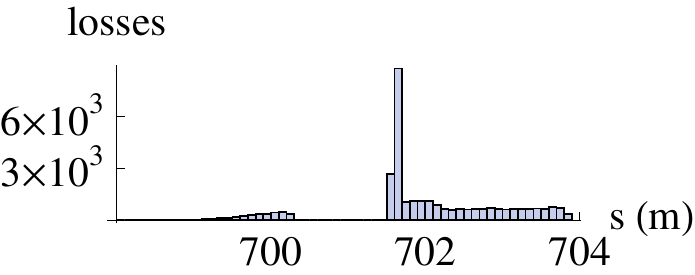}
  \includegraphics[width=4cm]{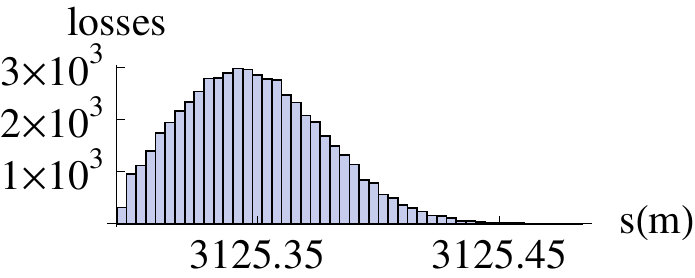}\\
  \caption{The distribution of $^{206}\mathrm{Pb}^{82+}$ losses after IP2. The two bottom plots show zooms on the largest peaks on the upper plot. Only the loss at $s\approx3125$~m risks to induce quenches, because of the small projected spot size.}
\label{fig:EMD2-lossmap}
\end{figure}

\begin{figure}
  \includegraphics[width=7cm]{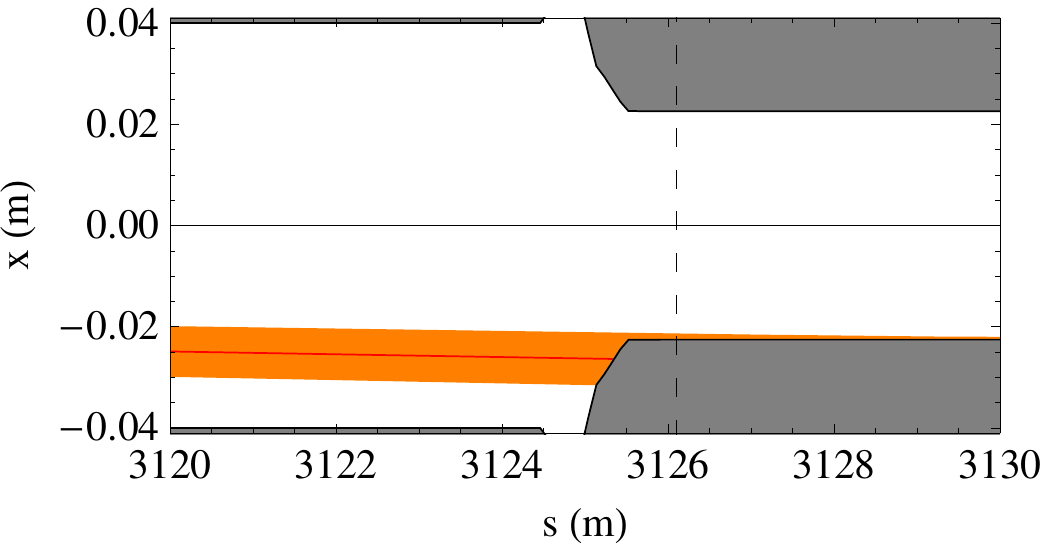}\\
  \caption{The impact  of the $6\sigma$ envelope of the
$^{206}\mathrm{Pb}^{82+}$ ions created by
  EMD2 at IP2 at an aperture restriction.
  The projected longitudinal rms spot size on the beam screen is only 2.9~cm.
The loss
  takes place in a drift section and the beginning of next magnet, a corrector, is indicated by a dashed vertical
  line.}\label{fig:emd2neutIP2}
\end{figure}

We used Eq.~(\ref{eq:x-at-impact}), with $\xbfpp$ replaced by the central EMD2 trajectory, to track particles from IP2 with this $\delta_p$-distribution and a phase space distribution given by Eq.~(\ref{eq:Pbfpp}). The resulting loss map is shown in Fig.~\ref{fig:EMD2-lossmap}. Around half of the losses is distributed over several locations, where there is no risk of quenches due to the small intensity and the wide spot sizes, while the other half is lost at $s\approx3125\,\mathrm{m}$ towards the end of a drift, 70~cm in front of an orbit corrector magnet, called MCBC. Since the losses occur at a place where the aperture is steeply decreasing, they are distributed over a very small longitudinal distance with a standard deviation of 2.9~cm. Therefore, the heat density per particle is correspondingly higher than in the case of BFPP. In Fig.~\ref{fig:emd2neutIP2} we show a close-up of the impact of the 6-$\sigma$ envelope of the EMD2 
beam from IP2.

The quench limit of the corrector is not well known. It is a function of the current and magnetic field, which are highly dependent on the closed orbit and the global orbit correction scheme. Furthermore, no network model of the corrector exists, so detailed simulations with \FLUKA\ and the network model were not carried out. Assuming its quench limit to be similar to the dipole's, we conclude that there is a risk of quenching from the small spot size. Since the beam is automatically dumped if the corrector quenches, this can not be allowed to happen.

However, it is possible to exclude this magnet from the global orbit correction. This means that it will not be operating during heavy ion runs, excluding quenches at the price of possible orbit distortions. Therefore losses from EMD2 are less serious than from BFPP, since a dipole is necessary for machine operation. During the first ion runs at 1.97~TeV/nucleon, the corrector is much less likely to quench than at top energy, for the reasons explained in Sec.~\ref{sec:operating-cond}.

Further losses from EMD2 take place 433.7~m downstream of IP1 and IP5, in a drift section only 30~cm in front of a main quadrupole.  In this case the projected rms beam sizes of the impinging ions are 30~cm for both IP1 and IP5.  In spite of this small size, the risk of quenches is small for several reasons.  The quench limit of a quadrupole is around 40\% higher than that of a dipole~\cite{bocian02} and the cross for EMD2 section is about 10~times smaller than for BFPP.  Furthermore, during the measurements described in Ref.~\cite{bocian02}, it was found that the tolerable heat load is approximately 50\% higher in the ends of the magnet, where the magnetic field is lower and the cooling more efficient.  However, to quantitatively determine the margin to quench, network model simulations should be carried out.

\section{MONITORING LOSSES} \label{sec:monitoring}

\begin{figure} \centering
\includegraphics[width=8.0cm]{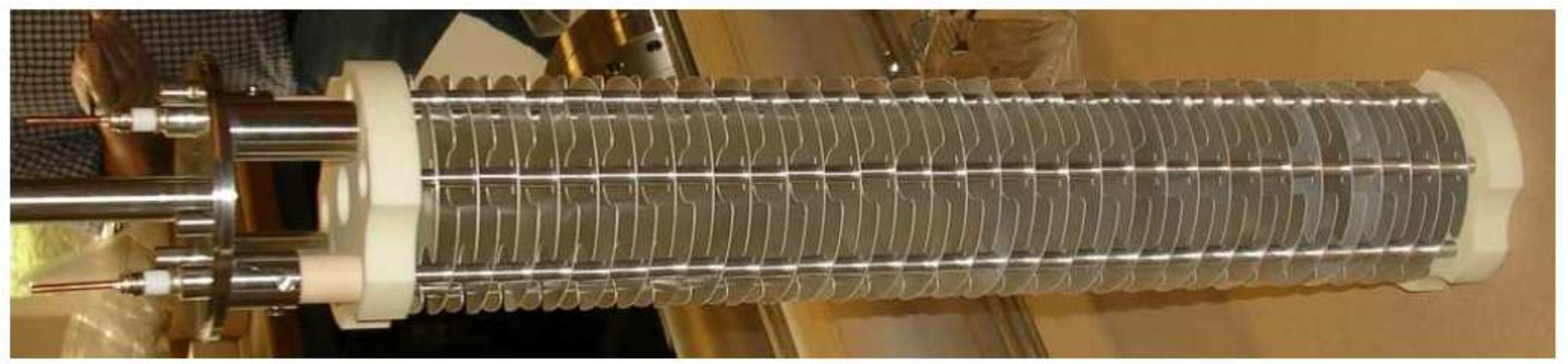}
\caption{Interior part of an ionization chamber used as a beam loss
monitor in the LHC. Taken from Ref.~\cite{holzer05}.}
\label{fig:blm-inner}
\end{figure}

Since beam losses caused by BFPP or EMD might induce quenches, it is vital to survey these losses while beams are colliding and to make sure that the beam is extracted quickly to a dump before a quench can occur. The LHC's beam-loss monitor (BLM) system has been designed to detect losses around the ring during proton operation~\cite{holzer05, holzer08a}. The LHC BLMs are 50~cm long ionization chambers filled with nitrogen that detect secondary charged shower particles outside the magnet cryostat. In order to minimize the drift time of the ions and electrons set free in the chamber, there is a stack of parallel aluminum plates inside the cylinder with alternating polarities and \mm{5} spacing.  A view of the interior part of an ionization chamber is shown in Fig.~\ref{fig:blm-inner}.

The locations of the BLMs and the thresholds for triggering the beam dump system have been determined for protons through simulations~\cite{gschwendtner02,holzer08b}. The BLM threshold signal should correspond to a a certain level of power deposition in the coil and  the ratio of these two quantities  could well be different depending on the type of particle lost and the showers it gives rise to. Accordingly, there are two important problems related to the monitoring of ion beam losses from the collisions:

\begin{enumerate}

    \item To determine the BLM threshold signals for Pb ion losses.

    \item To verify that the placement of BLMs provides adequate coverage of the loss patterns.

\end{enumerate}

To investigate the first problem, we simulated the development of the showers generated by particle losses, both from \pb\ ions and protons, in an LHC dipole magnet with \FLUKA; the geometry was as described in Sec.~\ref{sec:bfpp-shower} and Fig.~\ref{fig:lhc-MB-geo}. The BLMs are schematically modeled as thin rectangular iron boxes filled with nitrogen, placed outside the MB cryostat. This simplification is consistent with simulations done for protons~\cite{holzer05} and can be considered reasonable since we are primarily interested in the ratio between heat deposition from heavy-ion and proton losses rather than the absolute signal from the BLM.

For each particle species, a generic beam loss was represented by a ``pencil'' beam:  a monoenergetic beam of particles, all hitting the vacuum chamber at the same point (in the horizontal plane, 1~cm from the entrance of the magnet) at the same angle of incidence (0.5~mrad). An arbitrary loss can be modeled as a superposition of pencil beams. Simulations were done with \pb\ ions and protons at 2.76~TeV/nucleon and with protons at 7~TeV.

\begin{figure} \centering
  \includegraphics[width=7cm]{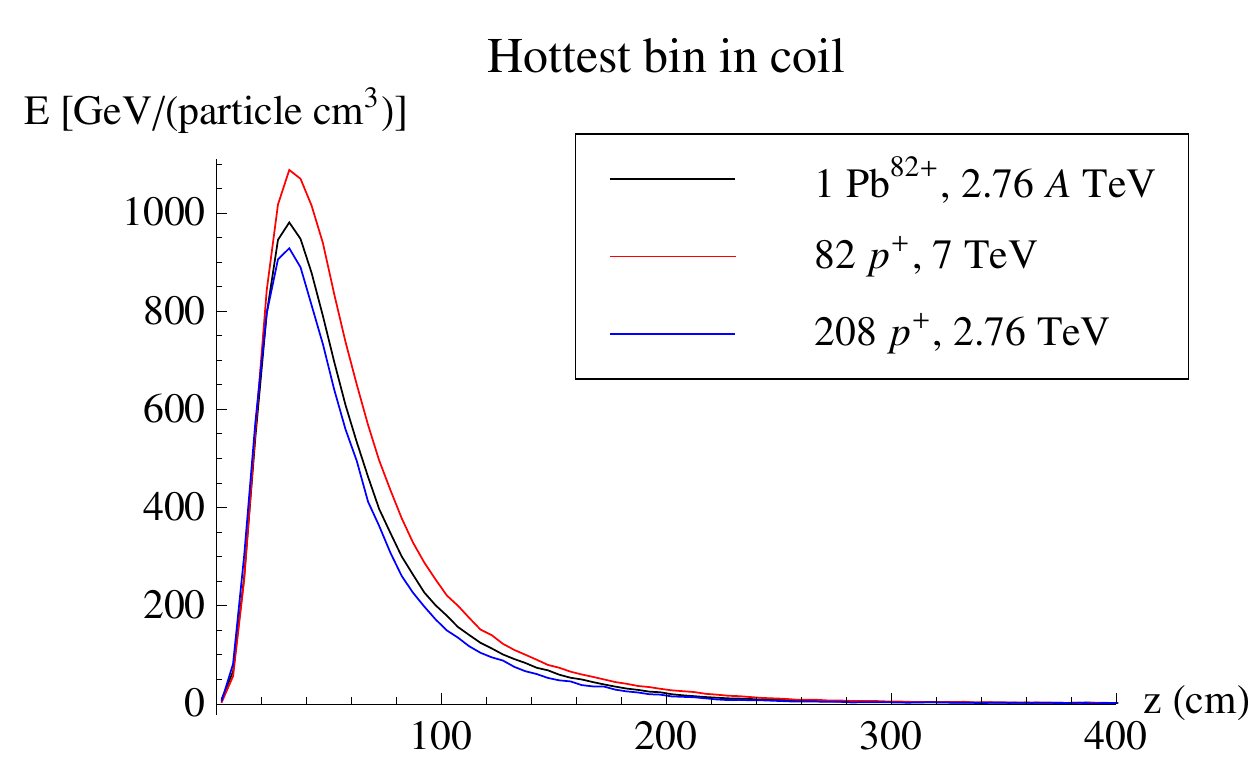}\\
  \includegraphics[width=7cm]{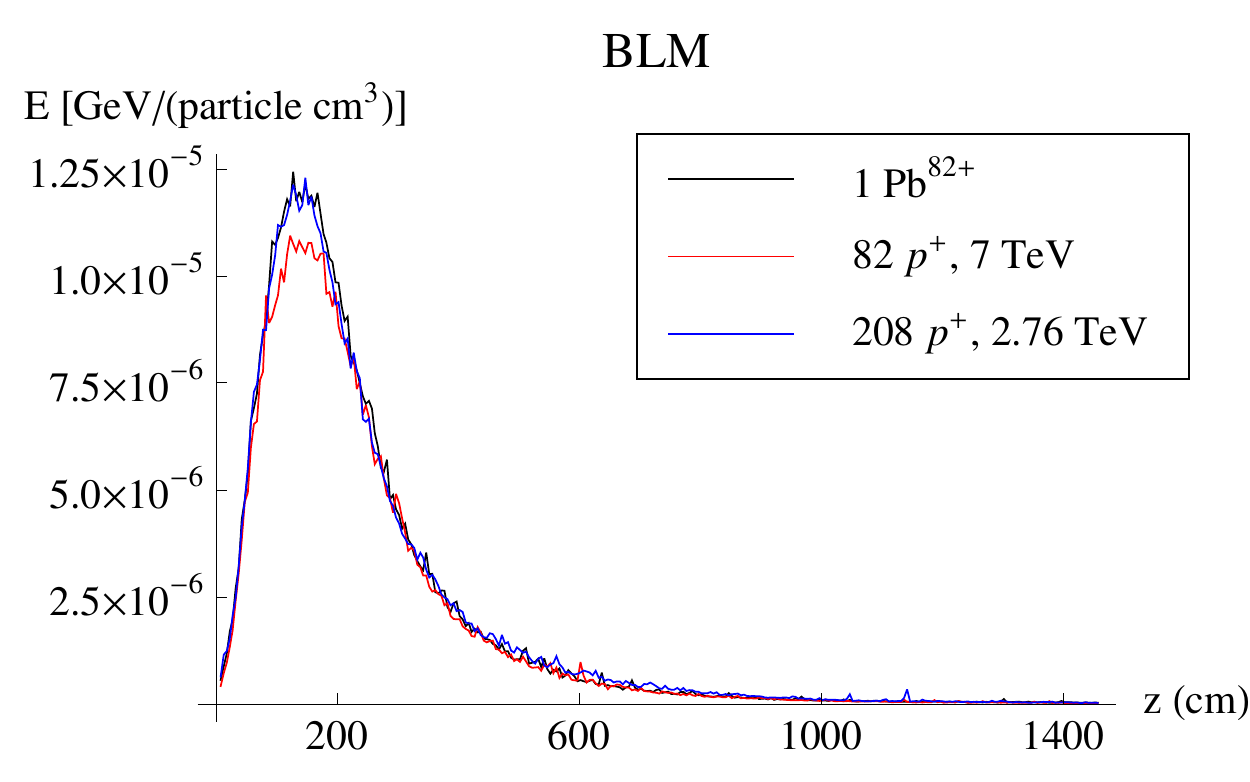}\\
  \caption{The energy deposition in the hottest wire in the coil and in the $\mathrm{N}_2$ gas inside the ionization chamber for Pb ions and protons.  The three cases are scaled to equivalent total energy of incoming particles.}\label{fig:edep-ratio}
\end{figure}

During the simulation, the energy deposition was scored in the beam screen, in the superconducting coil (using the mesh described in Sec.~\ref{sec:bfpp-shower}) and in the BLMs.  Since the BLM signal is proportional to the ionization energy loss in its gas volume and ionization is the dominant energy loss process for the low energy charged secondaries that emerge outside the cryostat, this is a fair approximation.  The resulting energy deposition profiles along the hottest superconducting cable in the MB coil and in the closest BLM are shown in Fig.~\ref{fig:edep-ratio}. The ratio between the heat deposited in the coils and the energy deposition in the BLMs is similar for \pb\ ions and protons at the same energy per nucleon, while 7~TeV protons have an even lower ratio between BLM signal and heating of the coil than 2.76 TeV/nucleon \pb\ ions.

\begin{figure} \centering
  \includegraphics[width=7cm]{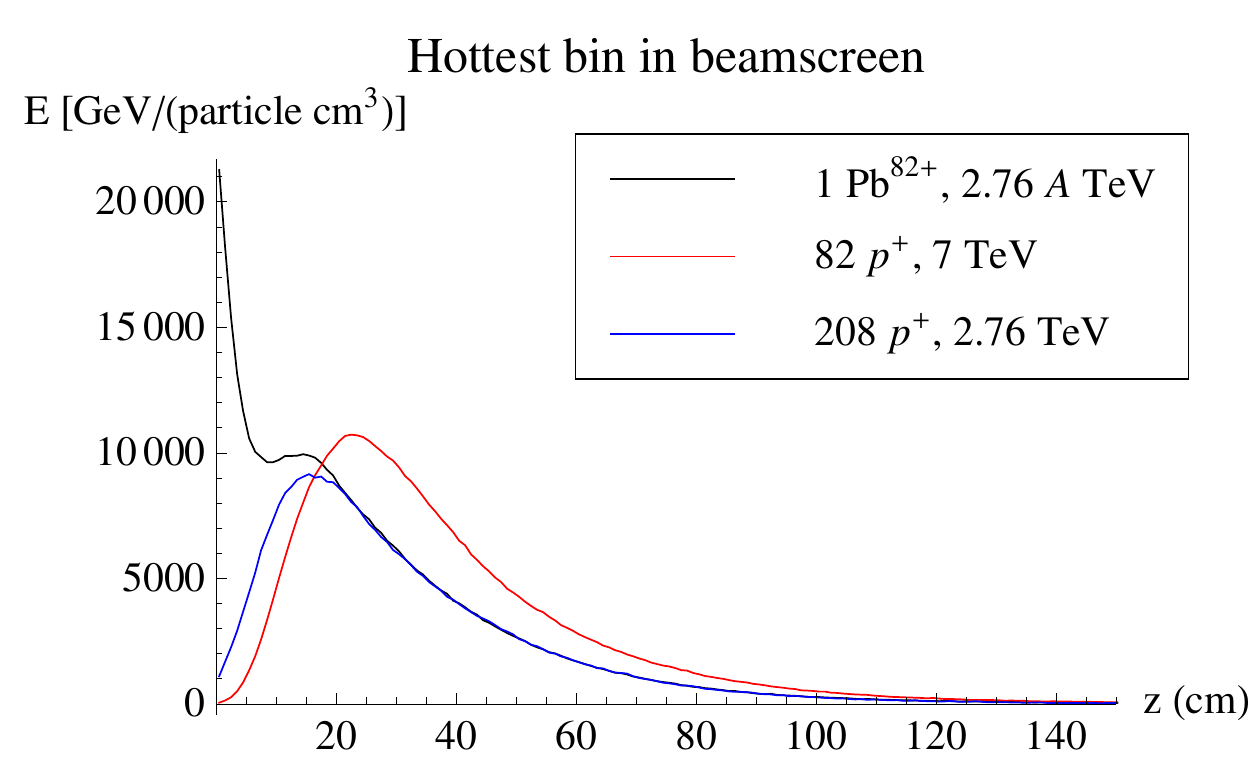}
  \caption{The energy deposition for the \pb\ ions and protons at
  LHC energy in the beam screen, as a function of the
  longitudinal coordinate in the magnet, averaged over
  $0.1\times 0.1\,\mathrm{mm}^2$ transversally and \mm{10}
  longitudinally.}
  \label{fig:edep-beam-screen}
\end{figure}

This similarity comes from the fact that the particles causing ionization of the gas in the BLMs are not the \pb\ ions and protons lost directly from the beam, but instead low-energy secondary particles, mainly electrons, created in the hadronic shower.  The energy deposition resulting from the hadronic shower is very similar for heavy ions and protons~\cite{agosteo01}, even though the nuclear interaction lengths are very different (0.8~cm for 2.76~TeV/nucleon \pb and 25~cm for 7~TeV protons on a Cu target according to \FLUKA\ simulations). This comes from the fact that when an ion traverses a target, the nucleus splits up into smaller fragments through electromagnetic dissociation and nuclear interactions in several steps. Once totally fragmented it gives rise to a similar shower profile as independent nucleons. The energy deposition from the first few interactions is far exceeded by deposition by the low energy secondary particles created later in the shower.

If the energy per nucleon increases, so do the cross sections, and more energy is deposited closer to the impact. This  explains why 82 protons at  7~TeV have a somewhat higher energy deposition in the coil than the 208 protons at 2.76~TeV.

The only location where a significant difference in energy deposition between \pb\ ions and protons can be seen is in the beginning of the shower, close to the central core.  Here the difference in ionization energy loss, given approximately by the well-known $Z_0^2$ dependence in the Bethe-Bloch formula~\cite{pdg}, is clearly visible and the ions cause a much higher energy deposition. Thanks  to the small impact angle, this effect is only visible in the beam screen, as shown in Fig.~\ref{fig:edep-beam-screen}.

Based on the similar ratio between energy deposition in the coil and in the BLMs outside the cryostat, we conclude that we can use the same beam dump thresholds for \pb\ ions and protons in the LHC.  This result applies to all mechanisms for ion beam losses in the LHC, not only those discussed in this paper.

Suitable positions for BLMs were determined from the optical studies illustrated in Fig.~\ref{fig:off-mom-env-2d}. Studies were performed both for the nominal parameters and a configuration known as the ``early ion scheme'' optics (which has a higher $\beta^*=1.0\,\mathrm{m}$, see Chap.~21 in Ref.~\cite{lhcdesignV1}) and for the two beams circulating in opposite directions.

Because of the uncertainty of the impact position, described in Sec.~\ref{sec:sensitivity}, and the fact that the loss peaks are narrow and localized, the  LHC's machine protection system needs a very tight coverage with BLMs. Based on the energy deposition profile in the BLM gas (see Fig.~\ref{fig:edep-ratio}) a 1.5~m spacing between chambers has been assumed for both beams to ensure full detection and localization of losses.

The BLMs foreseen for proton operation  are mounted on all quadrupoles in the arcs and dispersion suppressors~\cite{holzer05}. Extra chambers for  monitoring ion losses from BFPP have been requested in all loss locations that are not already covered. Since a fraction of the losses might escape the first loss location and instead hit the aperture close to the second peak of the dispersion, this position will also be monitored.

No extra BLMs are needed for losses from EMD2, since the loss locations are already covered by the default scheme.

\section{ALLEVIATION OF LOSSES} \label{sec:alleviation}

Ideally,   the optics and aperture of a heavy-ion collider would be such that the transformed ions stay within the acceptance of the ring and are cleaned by the collimators. But this may conflict with other design constraints like the overwhelming cost advantage of keeping the aperture small. In the case of the LHC, quenches caused by lost BFPP and EMD2 ions are likely and methods to avoid them have to be found. Several methods using specialized hardware are possible, the most promising one being the installation of extra collimators around each IP where the dispersion starts to rise in the cold part of the machine~\cite{lhcdesignV1,assmann09}. This could be a very efficient solution   but   requires extra hardware to be installed, as well as displacing some magnets to make room in the tunnel and cannot be done before the first heavy-ion physics runs.

For these runs at least, other methods to mitigate the effects of BFPP and EMD2 using existing hardware  will be needed. We cannot influence the quench limit of the dipole magnet, nor $\sigbfpp$, which leaves only $\lum$ and $W$ as the free parameters (see Eq.~(\ref{eq:heat-power})). It has been suggested that, for certain conditions on the filling time of the LHC, a higher integrated luminosity may be reached if the peak luminosity is  reduced~\cite{morsch01} by varying $\beta^*$   during a fill as a function of the remaining intensity. However this only pays off if the  single-bunch intensity can be raised. Otherwise the  only remaining   option is to reduce $W$ by varying the distribution of the impacting losses.

This might be achieved by manipulating the quadrupoles between the IP and the impact point to increase $\tilde{\beta}$ or $\tilde{d}$, spreading out the spot  of heat deposition. However,  strict requirements on the betatron phase advance around the LHC~\cite{aiba08}, matching conditions on       the periodic $\beta$ and dispersion functions  at the ends of the dispersion suppressors and constraints on magnet strengths limit the scope  of this approach. A gain in spot size of at best   20\% was achieved using this method.

\begin{figure}
  \includegraphics[width=8.5cm]{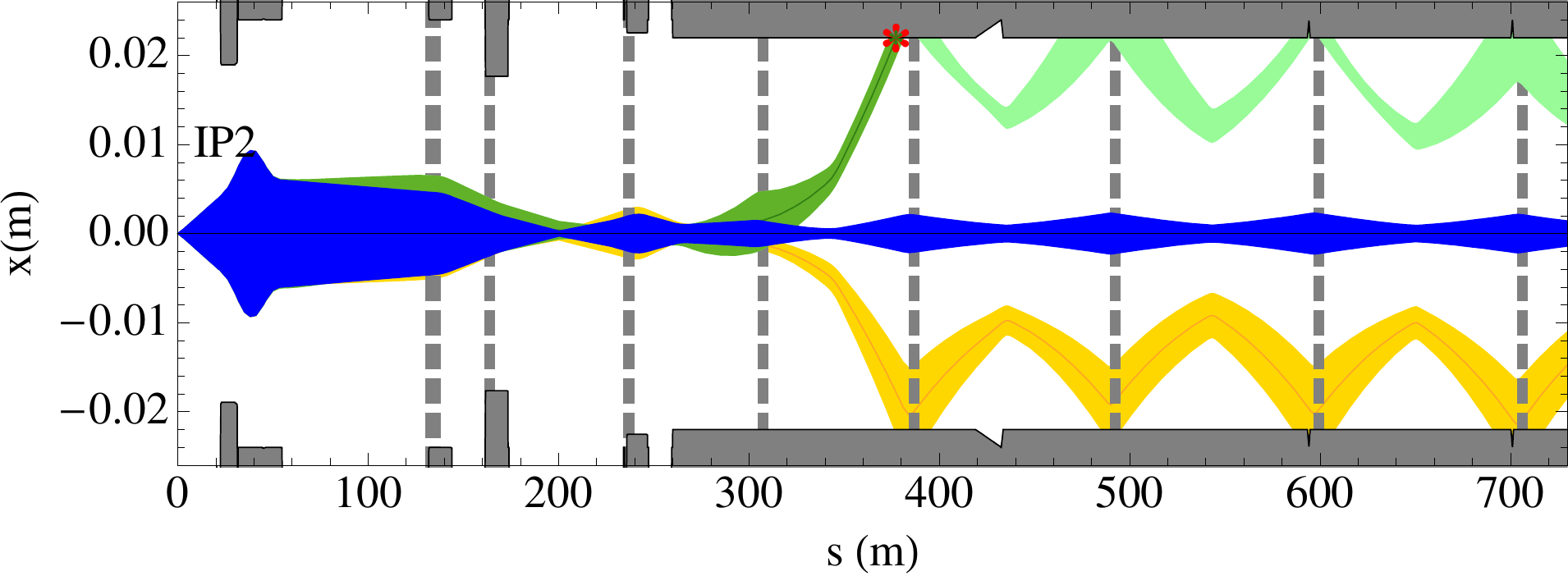}\\
  \includegraphics[width=8.5cm]{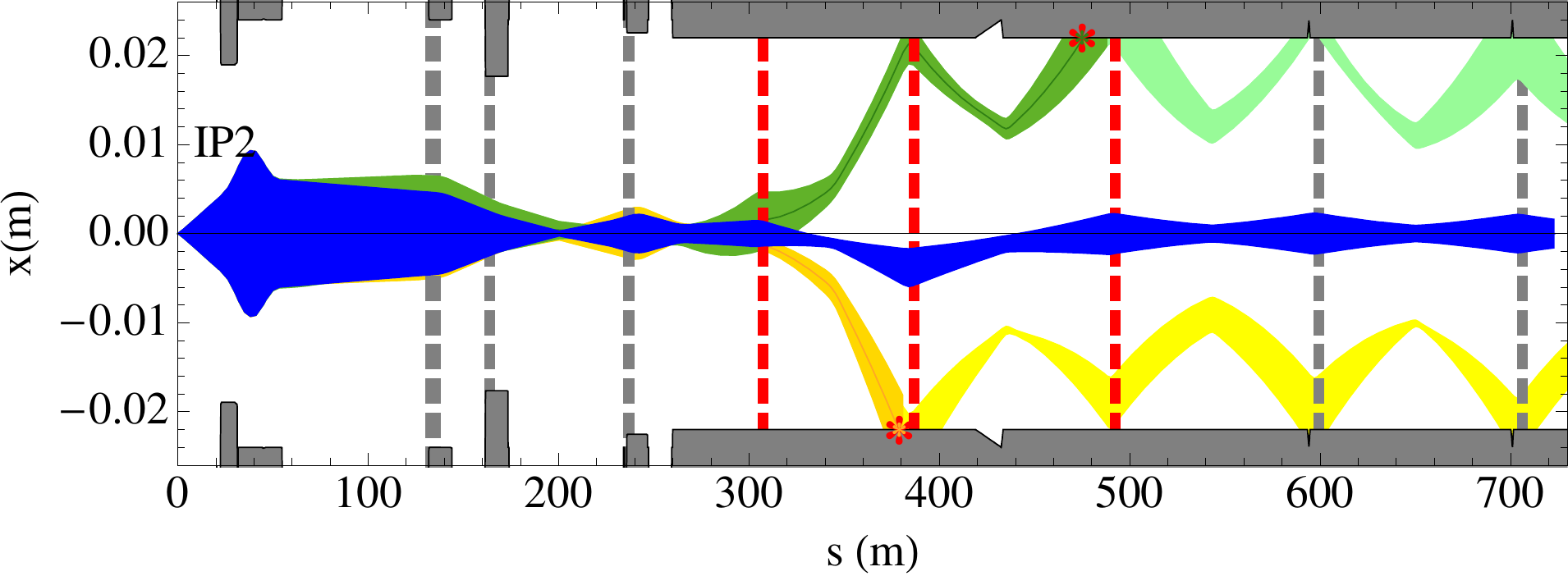}\\
  \includegraphics[width=8.5cm]{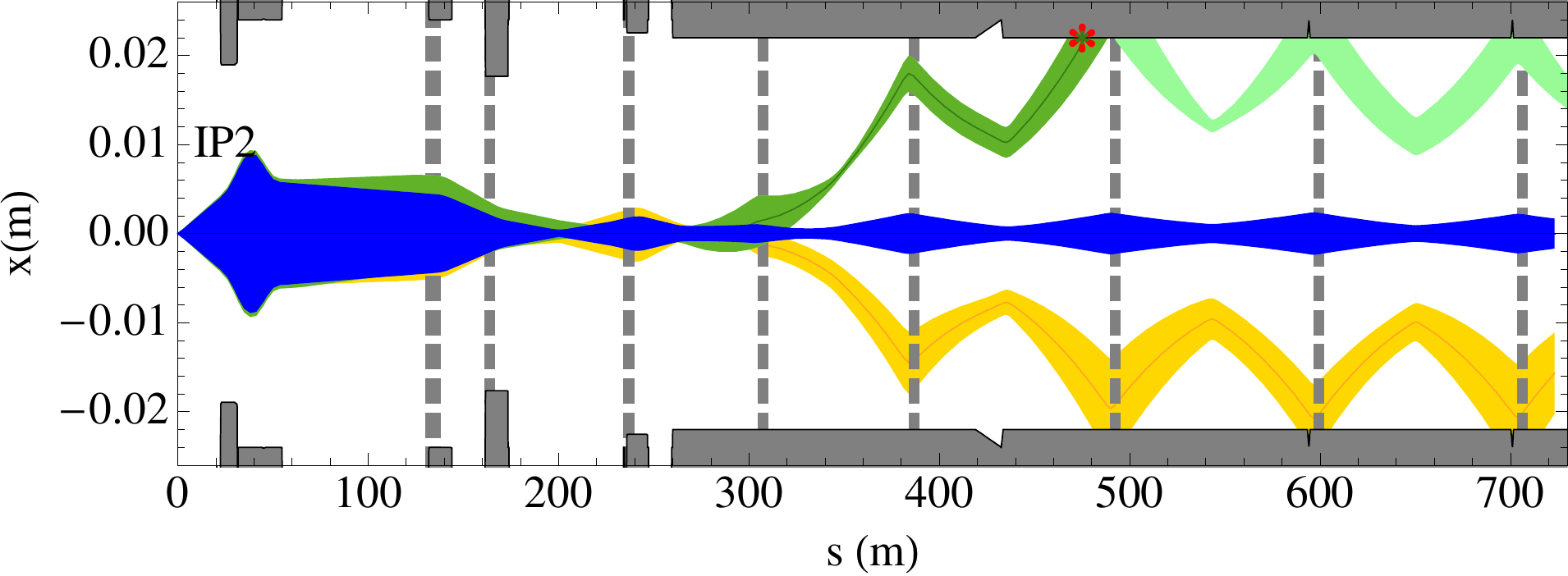}\\
   \caption{Horizontal physical aperture and 6$\sigma$~envelopes after IP2:
          with the usual flat orbit (top),
          with an orbit bump (middle), and
          with decreased dispersion (bottom).
          The beams are the main circulating \pb\ beam (blue),
          the $^{208}\mathrm{Pb}^{81+}$ BFPP beam (green), and
          the $^{206}\mathrm{Pb}^{82+}$ EMD2 (yellow).
          The vertical
          dashed lines show the locations of horizontal corrector magnets
          and those indicated in red are active
          (in this case, correctors called ``MCBC'' and ``MCB'').}\label{fig:BFPP-env-nom-bump}

\end{figure}

Another alleviation method might be to adjust the orbit or optics to move the impacts  to more favorable positions. The horizontal orbit of the particles affected by BFPP and EMD2 emerging from IP2 in a perfect lattice, as calculated by \MADX, is shown in the top part of Fig.~\ref{fig:BFPP-env-nom-bump} together with the main circulating beam. This is a close-up of the first 700~m in Fig.~\ref{fig:off-mom-env-2d}. The dispersive trajectories oscillate with $d$, and the BFPP particles are lost very close to its first maximum, which we call $s_1$, while the EMD2 beam continues further downstream.

The green BFPP envelope in Fig.~\ref{fig:BFPP-env-nom-bump}   is significantly wider at the second maximum $s_2$ of $d$, thanks to the larger value of $\tilde{\beta}$ (see Fig.~\ref{fig:off-mom-optics}). A localized closed orbit bump introduced around $s_1$  displaces $\xbfpp$ towards the center of the beam pipe and allows some of these particles to escape downstream to $s_2$. There, the lost ions are diluted over a larger volume decreasing the maximum power deposition.

At the same time, the envelope of the EMD2 ions moves towards the outside of the beam pipe at $s_1$  provoking losses there. On balance this is beneficial because the aperture is constant in the neighborhood of  $s_1$, so the losses are spread out much more than at the later loss point shown in Fig.~\ref{fig:emd2neutIP2}. This new loss position for the EMD2 ions is the same as for BFPP ions with a flat orbit  but since the EMD2 ion flux is almost a factor~10 lower than for BFPP, the heating at this position is kept low enough to avoid quenching the dipole (see Fig.~\ref{fig:iquench}).

The off-momentum orbit can be displaced by orbit correctors attached to the main quadrupoles in the LHC. The  orbit bump amplitude required was calculated to keep 95\% of the BFPP beam, with the beam size given by Eq.~(\ref{eq:BFPP-beam-size}),  inside the aperture at $s_1$. The resulting beam envelope with the bump active is shown in Fig.~\ref{fig:BFPP-env-nom-bump}.

The orbit of the circulating beam is displaced by up to  \mm{3.8} in this configuration but it is clear that the $6\sigma$~envelope is still far from the aperture. Apertures in the LHC are conventionally characterized using a quantity $n_1$, defined as the maximum acceptable primary collimator opening, in units of beam $\sigma$ that still provides a protection of the mechanical aperture against losses from the secondary beam halo~\cite{lhcdesignV1}; the complete definition incorporates a variety of tolerances that we shall not enumerate here. The bump reduces $n_1$  for the circulating beam,   from 30 to 21. This still provides sufficient aperture margin. The three corrector magnets  used to make the bump would operate at between 9\% and 26\% of their maximum strength, leaving a comfortable margin for their function in the global orbit correction. The $\beta$-beating due to the bump is around 0.3\%, which is acceptably small~\cite{lhcdesignV1}.

We have repeated the simulation chain of tracking, \FLUKA, and network model for the BFPP ions with the orbit bump included. The spot size at the new impact position, in another superconducting dipole magnet, is 220~cm and the maximum heat load from \FLUKA\ is $4.5\,\mathrm{mW}/\mathrm{cm^3}$, about a factor~3.5 less than without the bump. The result of network model simulation is included in Fig.~\ref{fig:iquench}. The expected heat distribution at $s_2$ with the orbit bump activated (indicated by a horizontal line) is predicted to be a factor 2.25 below the quench limit, still not completely safe.

Operationally, the bump amplitude will have to be fine-tuned around the theoretical   value to compensate the local closed-orbit and misalignments of the beam pipe. The BLMs placed around each loss location (as described in Sec.~\ref{sec:monitoring}) will be an essential tool to monitor how the BFPP losses move from the first to the second maximum of the dispersion.

The orbit bump will  be set up at low luminosity, in order to avoid potential quenches,  by means     of a van der Meer scan~\cite{vanderMeer68},  in which  the beam orbits are scanned \emph{vertically} across each other  at the IP. One could also imagine tuning the bump using a higher value of $\beta^*$ at the IP.  However, this is less reliable since the optical functions between the IP and the impact point will change with  $\beta^*$. Another option is to introduce the bump at lower beam intensity in an earlier fill and rely on reproducibility.

From the point of view of machine protection, we must consider the possibility that one of the three correctors used in the bump might quench,  leaving the bump open and possibly causing damage.  Two of them (of type MCBC) are directly connected to the beam abort system which would dump the beam immediately.  In case of a quench of the third corrector (of type MCB), no automatic beam dump occurs. The resulting global orbit distortion will, if it is small enough, be corrected by the orbit feedback system, in which case the BFPP beam may quench a magnet. If the distortion is too large, the resulting beam losses will trigger a beam abort via the BLMs (and possibly also the beam position monitors). In all cases, the quench protection system should protect the magnets from physical damage.

If moving the losses to another location with a larger spot size was insufficient, one might spread out the losses with several orbit bumps, which allow  fractions of the BFPP beam past each maximum of $d$. By tuning the orbit bumps the losses can be spread over $n$ dispersion maxima by $n-1$ orbit bumps using $n+1$ correctors in such a way that, ideally, a fraction $1/n$ is lost at each impact point. This decreases the maximum heating power in a single element by $1/n$ and can bring it below the quench limit if $n$ is made large enough. However, since BLMs have to be used when the bumps are tuned, the precision of the achieved loss distribution is limited.

To determine the required bump amplitudes $\Delta_i$ at each impact location $s_i$ we consider the initial phase space at the IP.  The particles lost at a location with horizontal aperture $A_i$ satisfy (for $i\in[1,n-1]$)
\begin{equation}
\label{eq:ap-ineq}
x_i-\Delta_i>A_i,
\end{equation}
where $x_i$ is given by Eq.~(\ref{eq:x-at-impact}) as a function of the initial conditions at the IP. These inequalities define regions $R_i$ in the initial phase space, which vary with $\Delta_i$.  The fraction $F_i$ of particles lost at $s_i$ is the integral distribution function over the phase space area outside the aperture limitation~(\ref{eq:ap-ineq}) and not outside any previous aperture limitation:
\begin{equation}
\label{eq:phase-space-integ}
F_i={\displaystyle \int_{R_i\cap \left ( R_1^c\cup R_2^c ... \cup R_{m-1}^c \right)} \Pbfpp(x_0,x_0')}
\Pdelta(\delta_p)\;dx_0\;dx_0'\;d\delta_p
\end{equation}
where $R^c$ denotes the complement of region $R$ and $\Pdelta$ is the assumed Gaussian distribution function of $\delta_p$. 

The $\Delta_i$ can then be determined by requiring
\begin{equation}
F_i=1/n, \, \forall \;i
\end{equation}
and solving Eqs.~(\ref{eq:phase-space-integ}) recursively, starting at $i=1$, which can be solved analytically to yield
\begin{equation}
\label{eq:delta1}
\Delta_1=\sqrt{2}\StandDevBfppOne
\,\erf^{-1}\left(\frac{n-2}{n}\right)+\xbfppI-A_1.
\end{equation}
Here $\StandDevBfppOne$ is given by Eq.~(\ref{eq:BFPP-beam-size}). Numerical integration and solution has to be used for higher values of $i$.

The method of   one or several orbit bumps   only works as long as the displacement of the closed orbit is kept within acceptable limits and the corrector magnets do not operate too close to their maximum strength. This is not the case at IP1 and IP5, where the required bump amplitude for BFPP particles to avoid the first maximum of $d$ is 8--9~mm. The method will not work there unless other changes are made to the optics.

Instead of an orbit bump, another method to move losses to $s_2$ would be to tune the quadrupoles after IP2 in order to decrease the dispersion at $s_1$. \MADX\ was used to rematch IP2 with all constraints mentioned in Ref.~\cite{aiba08} and beam envelopes for one solution are shown in the third plot of Fig.~\ref{fig:BFPP-env-nom-bump}. The  longitudinal size of the spot at $s_2$ is 179~cm, meaning a higher heating power  than with the orbit bump. So quenches cannot be excluded with this method either.

The required change in phase advance in the insertion is $0.05\times 2\pi$, which can be compensated in the insertion in IP4 containing the RF system~\cite{aiba08} to keep the overall tune in the machine constant. This method has the advantage of not displacing the circulating beam but does not reduce the far-downstream losses from  EMD2. Given the numerous other constraints on the LHC optics, it is  difficult to  envisage applying this method at IP1, IP2, and IP5 simultaneously.

\section{CONCLUSIONS}

Electromagnetic interactions such as bound-free pair production and electromagnetic dissociation in ultraperipheral nuclear collisions at the LHC, modify the  charge and mass of beam ions. These particles follow dispersive orbits until they are lost in locations determined  by the machine optics, aperture and the magnetic rigidity of the ions.  When sufficiently localized  they can heat superconducting magnetic elements enough to make them quench.

We have presented the first fully-integrated simulation chain of beam tracking, shower simulation and a thermal network model to evaluate the heat-flow and quench behavior of the superconducting coils  immersed in superfluid liquid helium. This simulation has been applied  to the most critical loss mechanism, BFPP, occurring  in the three heavy-ion collision points in the LHC. Heat deposition caused by \pbone\ ions in main dipoles downstream from IP1, IP2, and IP5 is expected to be 40\% above the quench limit, while $^{207}\mathrm{Pb}^{82+}$ from EMD1 stay within the acceptance of the arc and are cleaned by the collimation system. Furthermore, depending on its excitation level, there is some risk from quenches of a corrector magnet downstream of IP2 by  $^{206}\mathrm{Pb}^{82+}$ ions created through EMD2.

To avoid quenches,  an efficient beam-loss monitor system to detect these losses is needed. Shower simulations of the relations between the signals in the loss monitors and the energy deposition in superconducting coils have been carried out for \pb and proton beams. These demonstrate that, thanks to nuclear fragmentation, the beam abort  system can be set to trigger at the same signal level for both beams. Additional monitors  have been installed in critical locations for heavy-ion losses in the LHC.

Finally, we have investigated methods to alleviate the losses caused by BFPP and EMD2. We can move them to different locations in the machine, where the losses are spread out over a larger distance, lowering the energy deposition to around 45\% of the quench limit. Approaches using orbit correctors to create a local orbit bump or quadrupole tuning to decrease the dispersion  have been illustrated. These methods have the advantage of not requiring new hardware in the LHC but are not easily applied  to all interaction points and  do not provide sufficient safety margins against quenches. An efficient solution is likely to require new hardware such as additional collimators or masks in the dispersion suppressor sections.

\section{ACKNOWLEDGEMENTS}

We would like to thank A.~Ferrari for enlightening discussions, M.~Giovannozzi for helpful advice and the LHC misalignment data, M.~Magistris for the MB \FLUKA\ model, and M.~Aiba for providing rematching routines for the LHC optics. Other people we would like to thank for their help during the course of this work are R.W.~Assmann, B.~Dehning, J.B.~Jeanneret, S.~Russenschuck, B.~Schr\"{o}der, A.~Siemko, G.I.~Smirnov, D.~Tommasini, J.~Wenninger and S.M.~White. 


\end{document}